\documentclass[lettersize,journal]{IEEEtran}
\usepackage{amsmath,amsfonts}
\usepackage{algorithmic}
\usepackage{algorithm}
\usepackage{array}
\usepackage[caption=false,font=normalsize,labelfont=sf,textfont=sf]{subfig}
\usepackage{textcomp}
\usepackage{stfloats}
\usepackage{url}
\usepackage{verbatim}
\usepackage{graphicx}
\usepackage{cite}
\usepackage{makecell}

\begin{document}

\title{Enhancing EEG Signal Generation through a Hybrid Approach Integrating Reinforcement Learning and Diffusion Models}

\author{Yang An$^1$ $^2$, Yuhao Tong$^1$, Weikai Wang$^3$, Steven W. Su$^1$ $^2$*	

\thanks{$^{*}$ The co-responding author.}
\thanks{$^{1}$ Jinan Key Lab of Intelligent Rehabilitation Robotics, College of Artificial Intelligence and Big data for Medical Sciences, Shandong First Medical University
Shandong Academy of Medical Sciences.}
\thanks{$^{2}$ Faculty of Engineering and IT, University of Technology Sydney, NSW, 2007, Australia}
\thanks{$^{3}$ The Sixth Mobile Detachment of the First Mobile Corps of the Chinese People's Armed Police Force}

}

\markboth{Journal of \LaTeX\ Class Files,~Vol.~14, No.~8, August~2021}%
{Shell \MakeLowercase{\textit{et al.}}: A Sample Article Using IEEEtran.cls for IEEE Journals}


\maketitle

\begin{abstract}
  The present study introduces an innovative approach to the synthesis of Electroencephalogram (EEG) signals by integrating diffusion models with reinforcement learning. This integration addresses key challenges associated with traditional EEG data acquisition, including participant burden, privacy concerns, and the financial costs of obtaining high-fidelity clinical data. Our methodology enhances the generation of EEG signals with detailed temporal and spectral features, enriching the authenticity and diversity of synthetic datasets. The uniqueness of our approach lies in its capacity to concurrently model time-domain characteristics, such as waveform morphology, and frequency-domain features, including rhythmic brainwave patterns, within a cohesive generative framework. This is executed through the reinforcement learning model's autonomous selection of parameter update strategies, which steers the diffusion process to accurately reflect the complex dynamics inherent in EEG signals.

  We validate the efficacy of our approach using both the BCI Competition IV 2a dataset and a proprietary dataset, each collected under stringent experimental conditions. Our results indicate that the method preserves participant privacy by generating synthetic data that lacks biometric identifiers and concurrently improves the efficiency of model training by minimizing reliance on large annotated datasets. This research offers dual contributions: firstly, it advances EEG research by providing a novel tool for data augmentation and the advancement of machine learning algorithms; secondly, it enhances brain-computer interface technologies by offering a robust solution for training models on diverse and representative EEG datasets. Collectively, this study establishes a foundation for future investigations in neurological care and the development of tailored treatment protocols in neurorehabilitation.

\end{abstract}


\section{Introduction}
\IEEEPARstart{E}{lectroencephalography}
 is recognized as an essential neurodiagnostic tool, distinguished by its high temporal resolution in capturing brain electrical activity. The resulting signals are instrumental for elucidating a spectrum of neurological conditions and cognitive processes \cite{Ref8}. However, the clinical utility and research potential of EEG are often impeded by the complexities of data acquisition, which demand large and diverse datasets to ensure the robustness and generalizability of EEG-based analytical methodologies \cite{Ref4}.

The advent of machine learning has introduced sophisticated algorithms that can discern complex patterns within EEG data, thereby significantly enhancing diagnostic precision and paving the way for innovative applications in brain-computer interfacing \cite{Ref6}. Yet, these algorithms are predicated on the availability of extensive high-quality, annotated datasets for training—a challenge exacerbated by the constraints of data collection, including participant burden, privacy concerns, and the economic implications of clinical-grade data acquisition.

In response to these limitations, the scientific community has increasingly adopted generative models to synthetically generate EEG data. These models provide a viable solution by producing artificial datasets that closely replicate the statistical properties of authentic EEG recordings, offering a valuable adjunct to existing datasets, facilitating the validation of machine learning algorithms, and enabling the exploration of novel analytical techniques without the ethical and logistical considerations associated with additional patient data collection \cite{Ref10}.

This paper examines the role of generative models in the synthesis of EEG signals. We provide an in-depth analysis of the current challenges in EEG data acquisition and the theoretical underpinnings of generative modeling. We then detail our proposed methodology for generating synthetic EEG data, including a description of the model architecture, training protocols, and the potential applications of synthetic data in enhancing EEG signal analysis and machine learning applications. Our goal is to contribute to the advancement of EEG research and to expand its clinical and technological applications in the realms of neurological care and brain-computer interface development.

Within the domain of EEG data analysis, the enhancement of data quality and usability remains a pivotal area of focus. Recent advancements in research have illustrated that the implementation of cutting-edge data collection and augmentation methodologies can markedly enhance the reliability and utility of EEG data. Specifically, in the context of augmenting the efficacy of brain-computer interfaces predicated on Steady-State Visual Evoked Potentials (SSVEPs), an innovative strategy has been proffered in \cite{Ref7}. This technique employs StarGAN for the generation of synthetic SSVEP signals, thereby significantly enhancing the precision in classification and the rate of information transfer in SSVEP-BCIs, and concurrently diminishing the dependence on personalized calibration datasets.

Moreover, \cite{Ref11} delineates a sophisticated framework that dissects raw EEG data into three essential constituents: subject-specific feature, Motor-Imagery (MI)-task-specific feature, and noise. Through the amalgamation of discriminative and generative models, coupled with the application of a comprehensive array of fundamental training objectives and tactics, this framework accomplishes the proficient disentanglement of features, subsequently bolstering the generalizability of motor imagery BCIs. While this methodology has achieved outcomes that eclipse the performance metrics of existing optimal algorithms on three publicly accessible MI EEG datasets, it may require intricate network topologies and elaborate training regimens. In addition, it may confront the twin obstacles of computational resource consumption and the pursuit of temporal efficiency within practical deployment settings.

In the research documented in \cite{Ref17}, the authors have introduced an innovative hybrid neural network framework for the classification of EEG signals. This framework incorporates a specially tailored Generative Adversarial Network (GAN) to augment data and has developed a discriminative feature network aimed at recognizing motor imagery tasks. By applying filter bank techniques to the multiple sub-bands of motor imagery (MI) EEG, the framework extracts sparse Common Spatial Pattern (CSP) features and constructs a Convolutional Recurrent Neural Network (CRNN) classifier integrated with discriminative features. This methodology not only enhances the precision of cross-subject EEG motor imagery classification tasks but also bolsters the practical utility of BCI systems through the application of data and feature enhancement techniques. However, it is noted that the data generated by GANs may encounter limitations pertaining to quality and diversity.

Moreover, in \cite{Ref15}, investigators have integrated Conditional Variational Autoencoders (CVAE) with Generative Adversarial Networks to discern the latent representations of EEG signals. They synthesize EEG signals of specific categories by modulating the parameters of the generative model. This approach employs CVAE to refine the synthetic data, ensuring a closer resemblance to the class characteristics of authentic samples, while leveraging adversarial learning mechanisms to optimize the parameters of the generator, discriminator, and classifier. This tactic effectively surmounts the challenges associated with the application of deep learning to small-scale datasets through data augmentation, thereby markedly enhancing the accuracy of motor imagery task recognition. Nonetheless, the CVAE-GAN model may confront issues regarding stability and convergence during training, and it exhibits a heightened sensitivity to the selection and tuning of hyperparameters.

Deep learning technologies have showcased significant potential in the analysis of EEG signals. Within the scope of the research delineated in \cite{Ref16}, the authors have introduced a framework predicated on Deep Convolutional Generative Adversarial Networks (DCGANs), which is intended to generate synthetic EEG data. This strategy is designed to augment the training dataset and subsequently enhance the efficacy of BCI classifiers. The study encompassed an experimental paradigm that incorporated conditions of both focused and diverted attention, deploying an end-to-end deep convolutional neural network to address the classification task, and capitalizing on the DCGANs framework for data augmentation. Notably, under the diverted attention conditions, this methodology achieved a marked improvement in classification accuracy through data enhancement, thus substantiating the considerable merits of utilizing GANs for the augmentation of EEG data and the enhancement of BCI performance. Nonetheless, the training of GANs may encounter impediments such as mode collapse and instability, necessitating the adjustment of network parameters to produce high-quality data tailored to the requirements of diverse users.

In \cite{Ref9}, investigators proffered an innovative approach to EEG data augmentation predicated on a hybrid network model that integrates wide and deep architectures. This technique engenders an enriched feature set by amalgamating EEG characteristics from two subjects exhibiting the highest degree of similarity, consequently amplifying the veracity of EEG data. Furthermore, when juxtaposed with conventional deep learning methodologies, this approach has demonstrated pronounced advantages pertaining to training duration and memory expenditure.

Technological advancements in data augmentation for EEG signals have been notably substantial across various application realms, including emotion recognition and the prediction of epileptic seizures. These developments are instrumental in elevating the operational efficacy of BCI systems, as well as in refining the exactitude of emotional recognition and the precision of forecasting epileptic events. Within the scholarly work cited in \cite{Ref14}, investigators have proposed a task-driven methodology harnessing Conditional Wasserstein Generative Adversarial Networks (CWGAN), aimed at fabricating high-fidelity synthetic EEG data. This initiative is designed to fortify the efficacy of affective recognition endeavors. The methodology leverages GANs to engineer artificial datasets characterized by clear categorizations, closely mirroring the statistical patterns observed in authentic data. Furthermore, it integrates an affective classification algorithm as a task-specific network, thereby steering the developmental trajectory of the generative model. A pivotal advantage of this approach is its significant contribution to the enhancement of EEG-driven emotion recognition tasks and its capacity to produce elite-tier data annotated with discrete emotional identifiers through a task-driven framework. Nonetheless, the cultivation of GANs may demand substantial investment in terms of time and resources, and challenges such as mode collapse could potentially result in a dearth of variability within the synthesized datasets.

The research articulated in \cite{Ref20} introduces an EEG emotion recognition model that adeptly integrates attention mechanisms with generative adversarial networks. This strategy merges spatial and channel attention to standardize and amplify the raw EEG datasets, while also capitalizing on GANs to amplify the dataset volume. This innovative approach adeptly tackles the dearth of EEG signal data. The technique has attained a detection accuracy of 94.87\% when evaluated against the SEED dataset, and through the complementary interplay of attention mechanisms and GANs, it has successfully augmented the model's precision and stability.

The academic literature cited in \cite{Ref2} thoroughly examines the application of deep generative models to augment EEG data, with the aim of refining the capabilities of emotion recognition systems that rely on EEG data. Researchers have introduced three distinct data augmentation strategies that are based on Variational Autoencoders (VAE) and GAN. These strategies utilize both comprehensive and partial usage tactics to enhance the training datasets. By creating highly realistic EEG feature datasets, these approaches have notably improved the efficacy of emotion recognition models and have forged new paths for resolving the issue of limited EEG data availability. However, it is essential to conduct a quality assessment of the synthesized data to guarantee its beneficial impact on model training. The success of data augmentation is likely to depend on the performance of the chosen generative models and the diversity of the training data.

The study outlined in \cite{Ref13} presents a Graph Neural Network (GNN) approach that combines contrastive learning with GAN-based data augmentation, customized for emotion recognition tasks that are based on EEG signals. This technique uses GANs to augment data and further improves the quality of representations derived from EEG signals through the application of contrastive learning. It has demonstrated high accuracy in emotional classification on both the DEAP and MAHNOB-HCI datasets. However, the training process for GANs may require significant computational resources and time, and the success of contrastive learning may be contingent upon the use of carefully crafted data augmentation strategies. Within the work presented in \cite{Ref5}, the authors propose employing a Deep Convolutional Generative Adversarial Network (DCGAN) to generate synthetic EEG data. They utilize transfer learning to evaluate the effectiveness of four prominent deep learning models in predicting epileptic seizures. This method not only increases the accuracy of seizure prediction but also effectively addresses the challenge of a lack of high-quality data.

During the acquisition of EEG data, the process is inherently complex and labor-intensive, potentially leading to subject fatigue, which may adversely affect the integrity of the collected data. Consequently, the procurement of extensive EEG datasets poses a significant challenge in practical scenarios. Within the realm of brain-computer interfaces, the efficacy of machine learning algorithms is contingent upon a substantial corpus of training data, which is essential for discerning the distinct patterns and behavioral nuances among users. A scarcity of data can result in reduced model efficacy and recognition accuracy. Furthermore, the absence of exposure to a diverse array of data may impede the model's adaptability to novel users or settings, attributable to the distinctive EEG profiles of each individual. The representativeness and lack of bias in sample selection are critical to ensuring the reliability and generalizability of research outcomes. In the medical sector, particularly in applications pertaining to neurorehabilitation or assistive technologies, data insufficiency may restrict the development of tailored treatment protocols, consequently impacting therapeutic efficacy.

Despite the availability of diverse models for signal data generation, the creation of EEG data still faces significant challenges. Initially, EEG data, as an unstructured biological signal, demonstrates extremely complex intrinsic patterns that are not easily interpretable. Consequently, the generated signals require a degree of diversity and richness that is adequate to enhance the classification model's ability to identify and understand a range of signal patterns. The endeavor to replicate this complexity in order to produce EEG signals that closely resemble authentic brain activity is replete with difficulties.

Secondly, although generative adversarial networks have the potential to generate EEG signals, their effectiveness is contingent upon the generator's capacity to learn the underlying distribution of genuine data. In the case of EEG data, this latent space may be exceedingly expansive and enigmatic. Determining the optimal generative model architecture and parameter tuning to capture the dynamic characteristics of EEG signals presents a significant challenge. Furthermore, the training process of adversarial generative models can be unpredictable, especially when dealing with high-dimensional data, which may result in the generation of EEG signals with repetitive patterns or incoherence.

Lastly, EEG activity is influenced by a multitude of biological factors, including but not limited to brainwave frequency bands and synchronization mechanisms. Generative models must adhere to these physiological principles while striving to produce signals that are realistic. Numerous existing studies have generated signals that, although they bear a superficial resemblance to EEG signals, fail to fully capture the time-frequency and spatiotemporal characteristics of EEG, resulting in a lack of essential biometric information within the generated signals.

In response to the complex challenges inherent in the domain of EEG signal generation, 
this study presents an innovative diffusion-based methodology designed for the synthesis of EEG data. 
This methodology incorporates a suite of pivotal innovations, which are essential for advancing the field:

\begin{enumerate}
\item{{\bf Enhancement of Data Sets and Model Training:} The diffusion-based model presented herein offers substantial advantages for augmenting EEG data archives and optimizing deep learning model training protocols. In contexts where the collection of authentic EEG data is financially constrained and data availability is limited, this model effectively expands the current data corpus by generating synthetic EEG data, thereby adeptly overcoming the limitations of data acquisition. Additionally, it diminishes reliance on a single data source while concurrently ensuring or enhancing the veracity and diversity of training datasets for models.}
\item{{\bf Optimization via Reinforcement Learning:} In this investigation, we have employed a reinforcement learning algorithm to autonomously optimize the parameter update strategies for the generative model, facilitating a balanced enhancement in the model's capacity to discern both the temporal dynamics and spectral features of EEG signals. This reinforcement learning framework functions as a regulatory mechanism, effectively mitigating the loss of spectral information during the iterative improvement of temporal data, and simultaneously preventing the degradation of the network's inherent efficacy during the ongoing optimization of time-frequency network configurations. Moreover, this methodology is characterized by its ability to automatically extract and replicate key EEG signal characteristics, thereby obviating the need for extensive manual feature engineering interventions.}
\item{{\bf Fidelity and Diversity in Signal Generation:} The EEG signals synthesized by the methodology described herein closely mimic authentic signals in waveform morphology and adhere to the intrinsic characteristics of EEG signals across both the frequency and time-frequency domains. This approach is adept at replicating the complexity and variability observed in genuine EEG signals. The synthesized signals display a diversity in amplitude, frequency, and signal modality that is comparable to that of real data, thereby enhancing the model's capacity for generalization. Unlike adversarial generative models that are confined to replicating existing data, potentially hindering innovation and diversity, the methodology presented in this study promotes the generation of novel EEG samples. These samples, while not previously observed, are generated through an independent exploratory process, ensuring their rationality and plausibility.}
\item{{\bf Privacy Protection and Training Efficiency:} Since the synthetic EEG data generated in this study are devoid of biometric identifiers associated with actual subjects, their employment in research endeavors and model training exercises offers enhanced participant privacy. Additionally, the synthetic EEG data are compatible with semi-supervised and self-supervised learning frameworks, thereby reducing the necessity for extensive annotated datasets and augmenting the efficacy of model training protocols.}
\end{enumerate}

\section{Methodology}

\subsection{Block diagram}
\begin{figure}[!htb]
  \centering
  \includegraphics[width=3.5in]{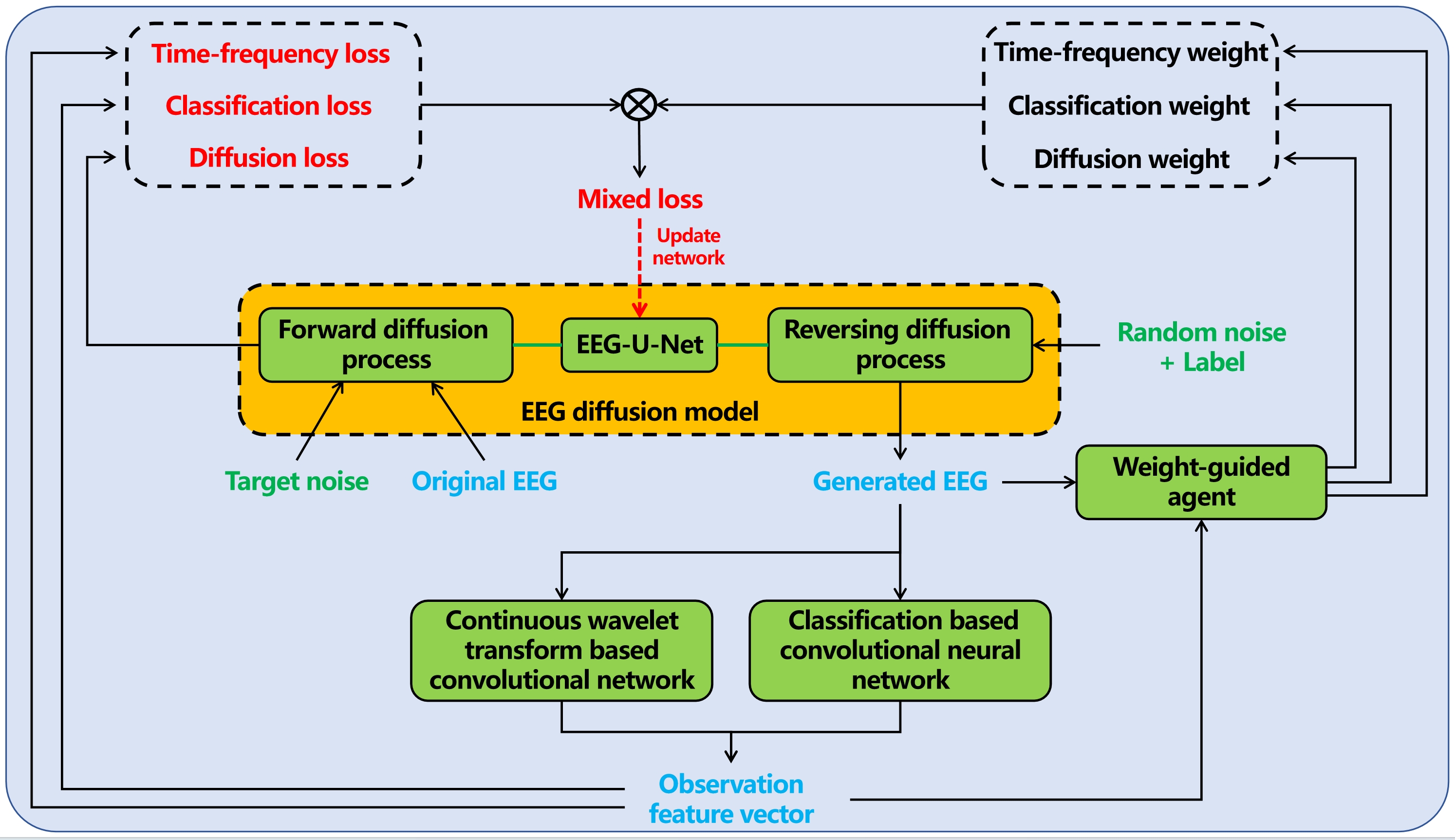}
  \caption{Block diagram of the actor-critic based EEG diffusion system.}
  \label{Block_diagram}
\end{figure}

The block diagram depicted in Fig.\ref{Block_diagram} elucidates the actor-critic based EEG diffusion system, which encompasses four core components: the EEG diffusion model, the continuous wavelet transform-based convolutional network, the classification-based convolutional neural network, and the weight-guided agent. Initially, the EEG diffusion model integrates both a forward and a reverse diffusion mechanism. During the forward diffusion process, the input comprises the target noise and the original EEG signal, culminating in the generation of a diffusion loss as the output. In contrast, the reverse diffusion process ingests random noise and the desired signal class as inputs, culminating in the synthesis of the EEG signal. Subsequently, the synthesized EEG signal is independently introduced into the continuous wavelet transform-based convolutional network and the classification-based convolutional neural network to extract the time-frequency feature vector and the classification feature vector, respectively. These vectors facilitate the calculation of the time-frequency loss and the classification loss. Thereafter, the aforementioned feature vectors are employed as observations and concatenated to generate EEG signals. These signals are subsequently supplied to the weight-guided agent to ascertain the weight allocated to each loss component. Ultimately, the loss weight is amalgamated with the time-frequency loss, classification loss, and diffusion loss to derive the composite mixed loss. This loss function serves as the foundation for the optimization of the parameters of the EEG-U-Net within the EEG diffusion model. This process ensures that the signals produced by the model encapsulate EEG information across the temporal, spatial, and frequency domains in tandem, with discernible classification distinctions among diverse categories.

\subsection{EEG diffusion model}

\subsubsection{Forward diffusion process}
\begin{figure}[!htb]
  \centering
  \includegraphics[width=3.5in]{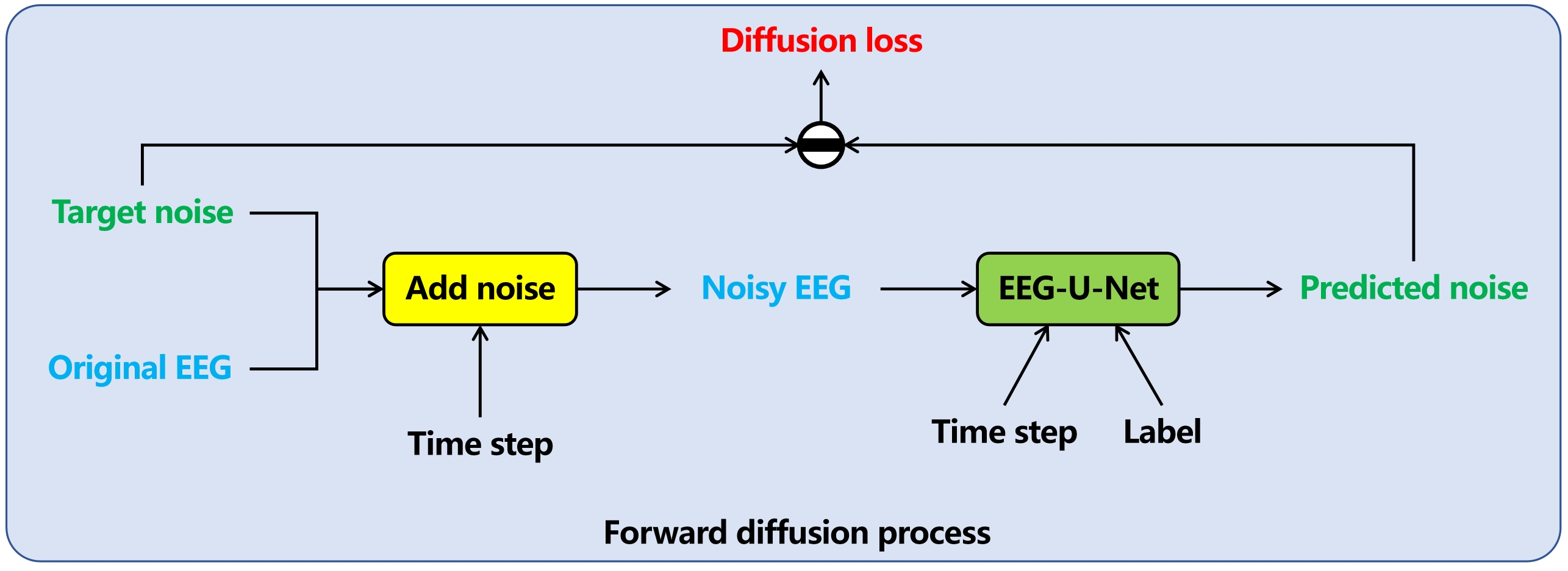}
  \caption{The forward diffusion process of EEG diffusion model.}
  \label{Forward}
\end{figure}

The EEG diffusion model encompasses two distinct procedural phases: the forward diffusion process and the reverse diffusion process. In the context of the forward diffusion process, as specifically illustrated in Fig.\ref{Forward}, there exist two fundamental elements: the initial incorporation of noise and the subsequent employment of the EEG-U-Net architectural framework for the anticipation of noise patterns.

Initially, the forward diffusion process systematically integrates Gaussian noise into the pristine EEG signal, progressively obfuscating it until the ultimate output is imperceptibly distinct from mere stochastic noise. At each temporal increment, Gaussian noise is superimposed employing the following equation:

\begin{equation}
  \label{Forward diffusion process}
  X_t = \sqrt{\prod_{i=1}^{t}(1-\beta_i)}X_0 + \sqrt{1-\prod_{i=1}^{t}(1-\beta_i)}\varepsilon 
\end{equation}

where $X_t$ is the noisy EEG; $X_0$ is the original EEG; $t$ is the time step; $\beta_i$ is the variance schedule. The variance schedule defines the noise level. 

Subsequently, the noisy EEG data, time steps, and target labels are introduced into the EEG-U-Net with the aim of noise prediction. The goal is to ascertain that the EEG-U-Net can accurately identify and replicate the noise introduced by the noise model, adhering to the temporal sequence. Consequently, the diffusion loss function of the EEG diffusion model is formulated from the squared discrepancy between the targeted noise and the noise output estimated by the model.

\subsubsection{Reverse diffusion process}
\begin{figure}[!htb]
  \centering
  \includegraphics[width=3.5in]{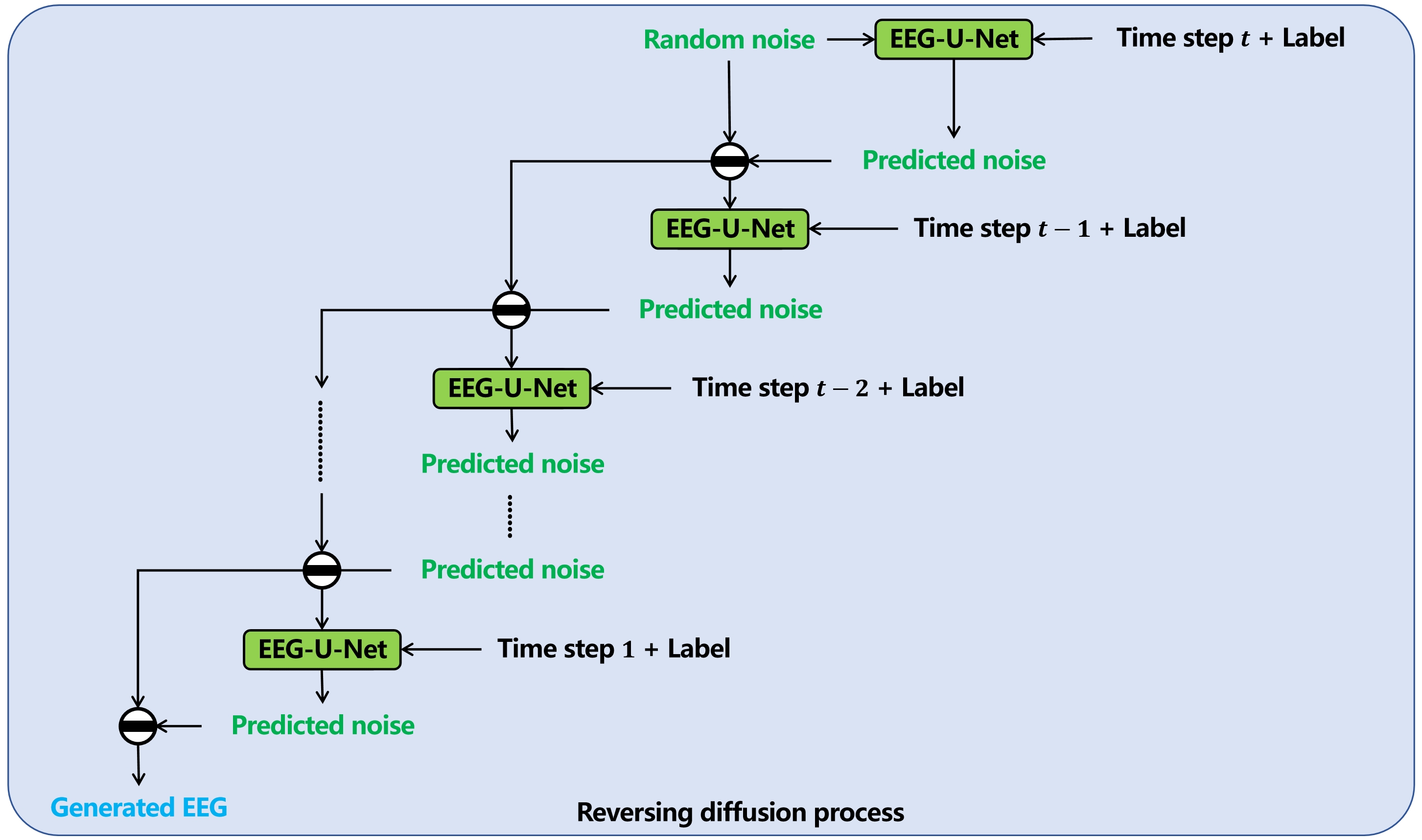}
  \caption{The backward diffusion process of EEG diffusion model.}
  \label{Backward}
\end{figure}

In the reverse diffusion process, as delineated in Fig.\ref{Backward}, Gaussian random noise is initially appointed as the EEG signal slated for reconstruction. This noise constitutes the commencement of our methodology. Subsequently, the EEG data, now adulterated with noise, in conjunction with the pertinent temporal information and categorical labels, are ingested by the EEG-U-Net to prognosticate the imposed noise. The excision of the predictive noise from the pristine synthetic EEG signal is actualized by the implementation of the ensuing formula:

\begin{equation}
  \label{Backward diffusion process}
  \begin{aligned} 
    Y_{t-1} = \frac{1}{\sqrt{1-\beta_t}}(Y_t-\frac{\beta_t}{\sqrt{1-\prod_{i=1}^{t}(1-\beta_i)}}\varepsilon_{pred})\\
    + \sqrt{\frac{1-\prod_{i=1}^{t-1}(1-\beta_i)}{1-\prod_{i=1}^{t}(1-\beta_i)}\beta_t}z \\
  \end{aligned}
\end{equation}

where $Y_t$ is the initial generated EEG signal; $\varepsilon_{pred}$ is the predicted added noise obtained from EEG-U-Net; $z$ is a Gaussian noise. 
At each iterative step, decrement the time step and reiteratedly perform the reverse diffusion process until the time step is reduced to unity. Ultimately, this procedure yields the variable $Y_1$ as the culminating generated signal.

\subsubsection{EEG-U-Net}
\begin{figure*}[!htb]
  \centering
  \includegraphics[width=7in]{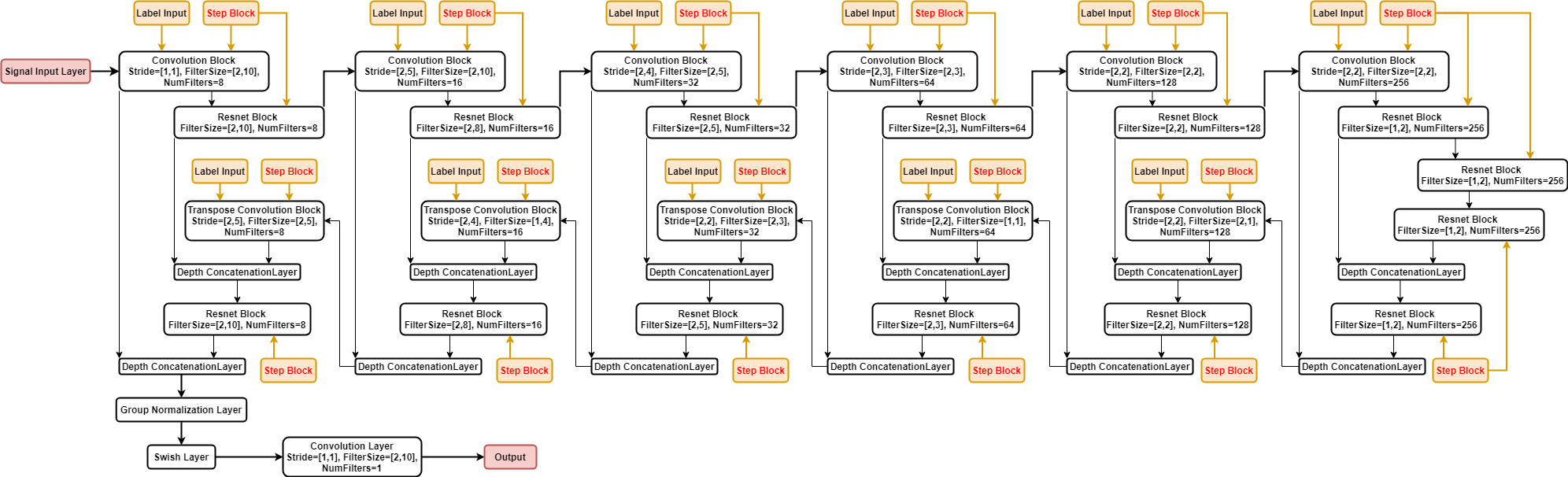}
  \caption{The architecture of EEG-U-Net.}
  \label{U_Net}
\end{figure*}

\begin{figure}[!htb]
  \centering
  \includegraphics[width=1.7in]{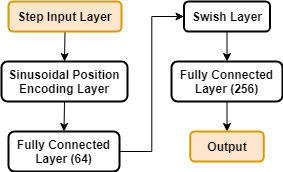}
  \caption{The step blocks of EEG-U-Net.}
  \label{StepBlock}
\end{figure}

\begin{figure*}[!htb]
  \centering
  \subfloat[]{\includegraphics[width=2.3in]{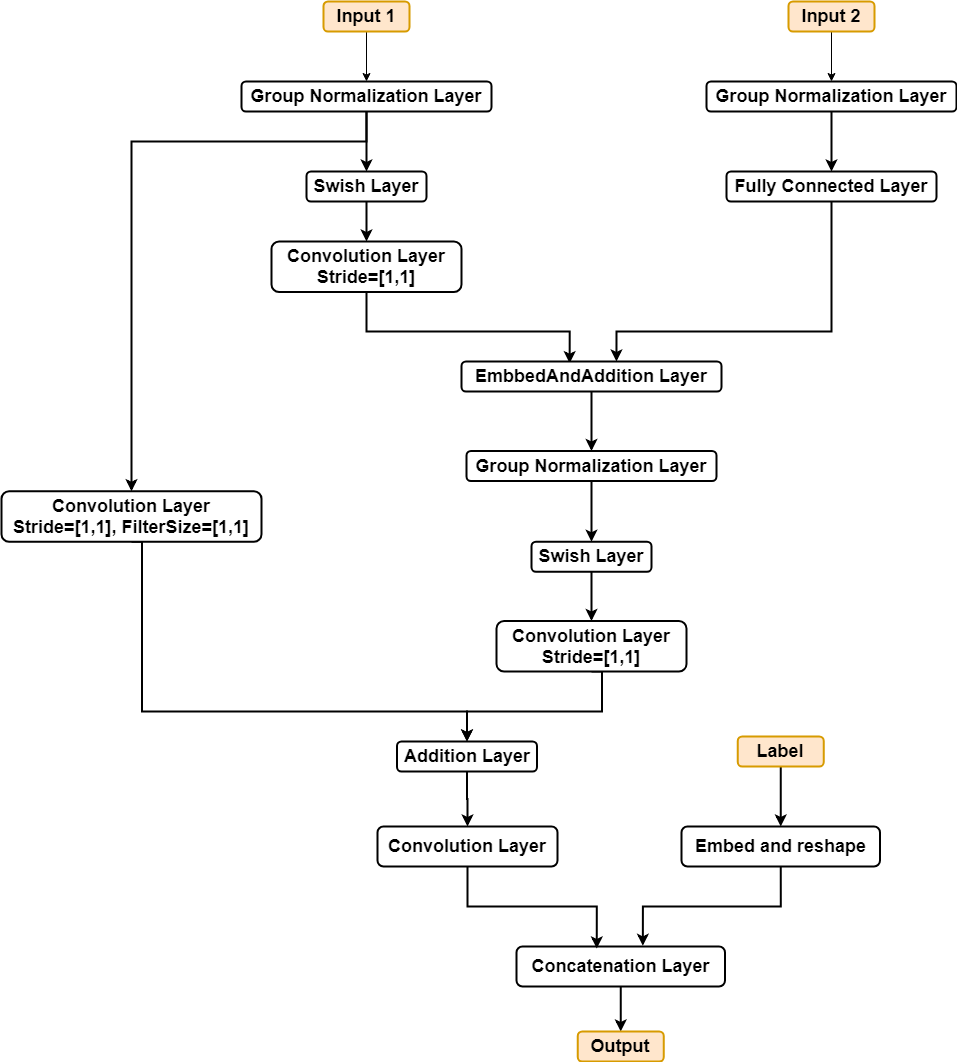}
  \label{ConBlock}}
  \hfil
  \subfloat[]{\includegraphics[width=2.3in]{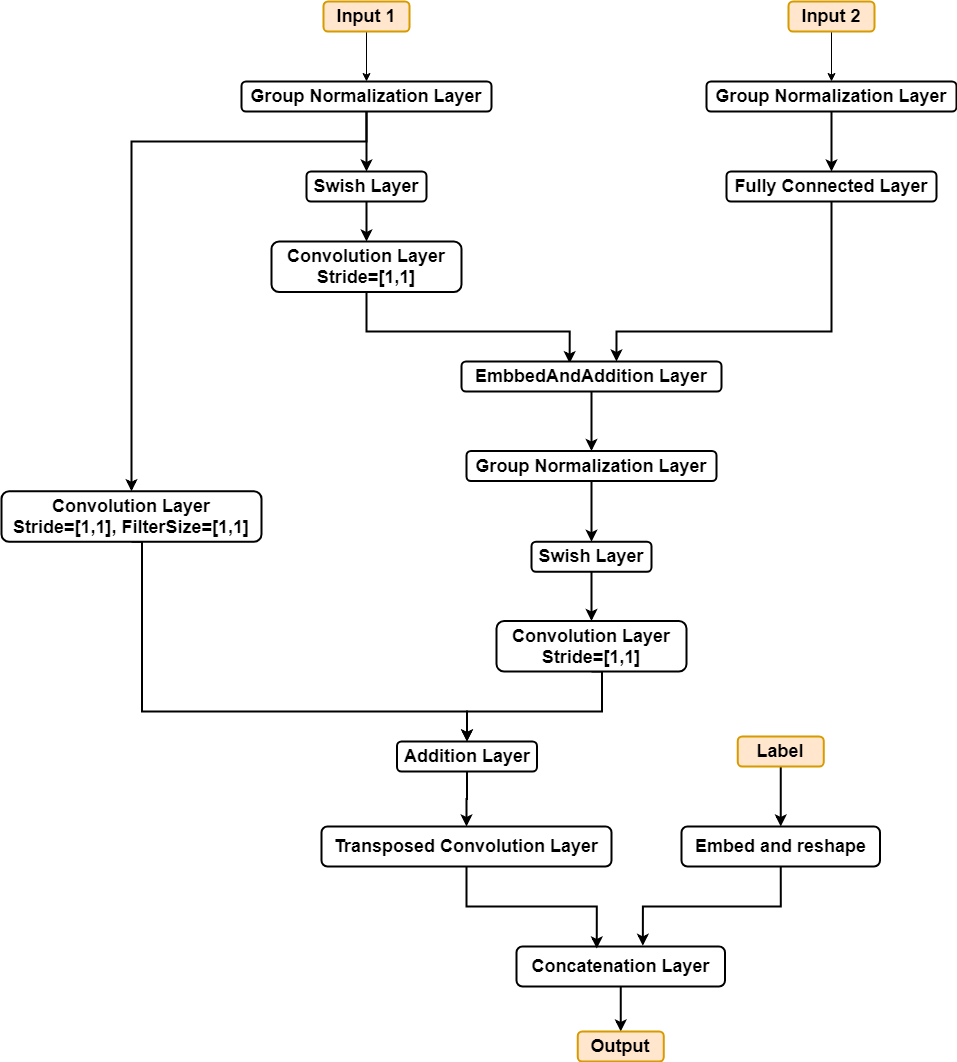}
  \label{TransConBlock}}
  \hfil
  \subfloat[]{\includegraphics[width=2.3in]{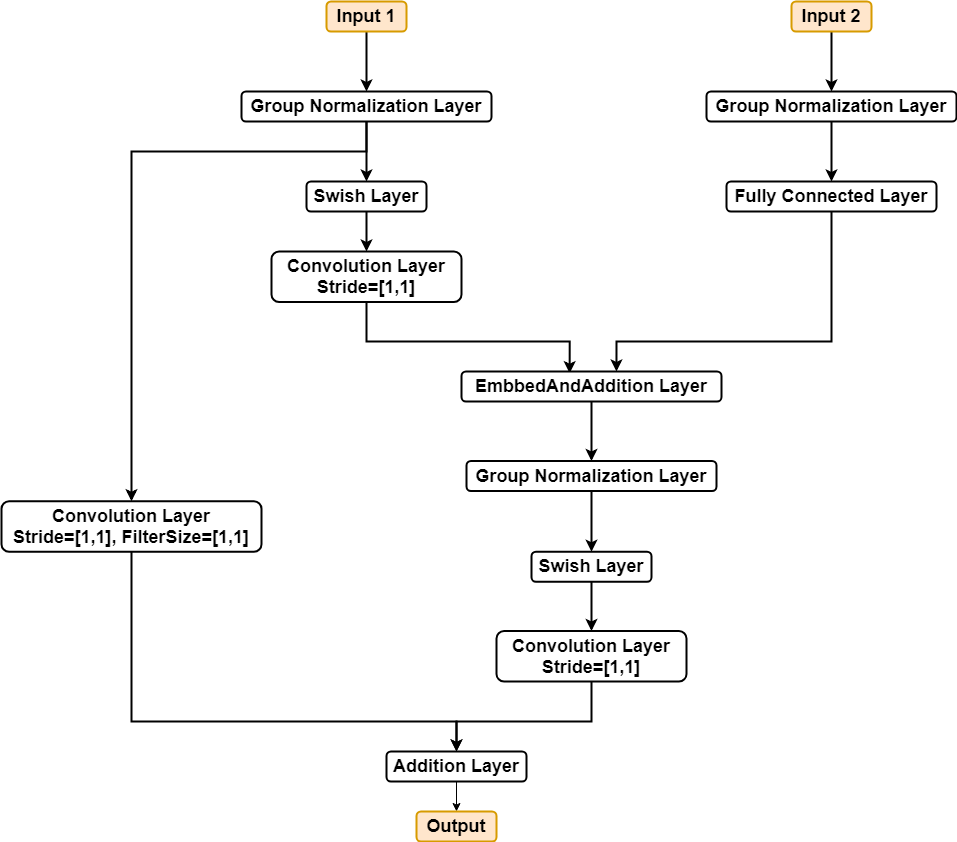}
  \label{ResBlock}}

  \caption{The architecture of important network blocks in EEG-U-Net. (a) Convolution block. (b) Transposed convolution block. (c) Residual block.}
  \label{NetBlock}
\end{figure*}

The fundamental goal of the EEG-U-Net is to effectively extract noise information from signals by utilizing the features learned from the signals. This robust network architecture employs a conventional encoder-decoder framework. Upon the encoder component's successful extraction of the pertinent features, the decoder component then maps these features back into the spatial domain of the original signal, progressively reconstructing it. Throughout this process, normalization is effected through the use of multiple global normalization layers, ensuring consistent scaling of features. 
The architecture of the EEG-U-Net is presented in Fig.\ref{U_Net}. 
The network consists of step blocks (as shown in Fig.\ref{StepBlock}), convolution blocks (as displayed in Fig.\ref{ConBlock}), 
transposed convolution blocks (as depicted in Fig.\ref{TransConBlock}), and residual blocks (as portrayed in Fig.\ref{ResBlock}).

Residual connections act as an architectural element that facilitates the direct conveyance of data from the lower tiers of the network to its upper echelons. This mechanism is instrumental in alleviating the gradient vanishing issue often encountered during the training regimen of deep neural networks, thereby empowering the network to assimilate more complex functions. Within the ResUnet architectural framework, residual connections can be effortlessly interwoven into each downsampling and upsampling phase. Dimensionality alignment is executed through convolution operations that utilize a kernel size of 1, consequently enhancing both the training efficacy and the overall performance of the model. The integration of residual connections significantly augments the capacity of the U-Net to address sophisticated and nuanced deep learning challenges, especially within the realm of EEG signals that demand high-precision feature extraction. This composite model, which amalgamates the symmetrical architecture of the U-Net with the profound learning capabilities of the residual network, adeptly manages extensive and complex signal data while sustaining a superior level of segmentation accuracy.

The network integrates skip connections, facilitating the concatenation of the feature vector from the corresponding layer of the encoder with that of the decoder. This merging process substantially augments the depth of feature information. The capture of contextual information is essential for the generation of EEG signals with superior quality and fidelity, as it permits the model to synthesize both local and global data. The U-Net structure's downsampling pathway effectively distills feature representations from the input noise and conveys these pivotal features to the upsampling pathway for signal synthesis. This proficient feature transfer mechanism ensures that the model retains crucial signal details and structures throughout the generative process, thereby enhancing the integrity of the generated signal.

Furthermore, we have adopted the Swish function as the activation function, a nonlinear modification that harmoniously combines the characteristics of the Sigmoid and ReLU functions. This ensures a nonzero gradient even when the input is negative, thus circumventing the potential issue of neurons becoming nonfunctional due to the utilization of ReLU. By embedding the Swish activation layer into the U-Net architecture, we amplify the model's capacity for nonlinear expression, which fosters advanced learning of complex data distributions.

\subsection{Continuous wavelet transform and classification based convolutional networks}

\begin{figure*}[!htb]
  \centering
  \subfloat[]{\includegraphics[width=2.5in]{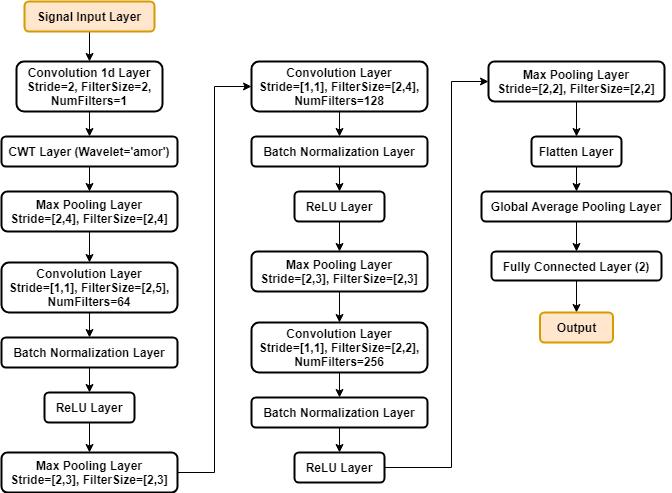}
  \label{CWTNet}}
  \hfil
  \subfloat[]{\includegraphics[width=2.5in]{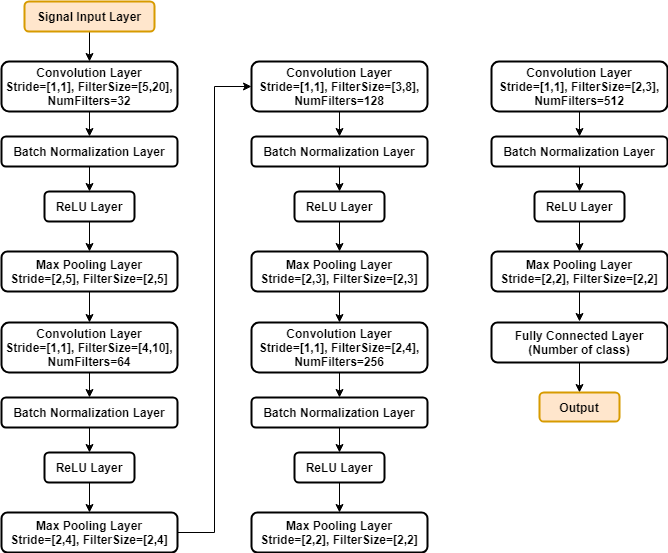}
  \label{ClassNet}}

  \caption{The architecture of continuous wavelet transform based convolutional networks and classification based convolutional networks. (a) Wavelet network. (b) Classification network.}
  \label{CWTNet and ClassNet}
\end{figure*}

A wavelet network represents a pioneering computational paradigm that amalgamates wavelet transformation with the technological underpinnings of convolutional neural networks. This model leverages the unique local and multi-scale attributes of wavelet functions, thereby affording enhanced performance in the realms of signal processing and feature extraction. It executes multi-resolution analysis, facilitating the efficient acquisition of localized signal nuances across a spectrum of scales. Within the purview of brain signal processing, the compact wavelet network demonstrates particular proficiency in identifying characteristics that span both temporal and frequency dimensions. The architectural design of the wavelet network is tailored to more effectively retain local features throughout the network. Owing to its mechanism of parameter sharing, it facilitates the acquisition of universal features across diverse scales, thereby diminishing complexity and bolstering computational efficiency. The schematic representation of the network's architecture is delineated in Fig.\ref{CWTNet and ClassNet}.

\subsection{Weight-guided agent}

\subsubsection{Structure of agent}
\begin{figure*}[!htb]
  \centering
  \subfloat[]{\includegraphics[width=3.5in]{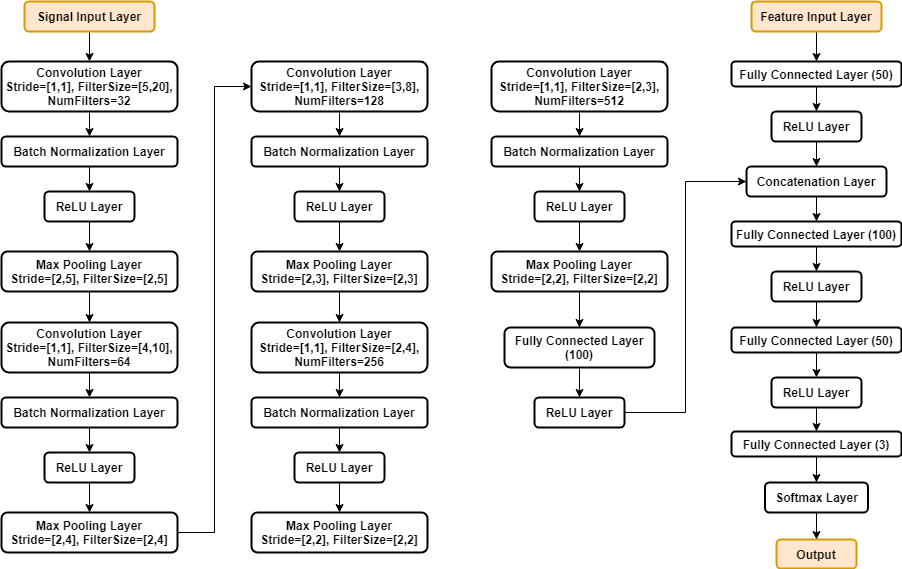}%
  \label{ActorNet}}
  \hfil
  \subfloat[]{\includegraphics[width=3.5in]{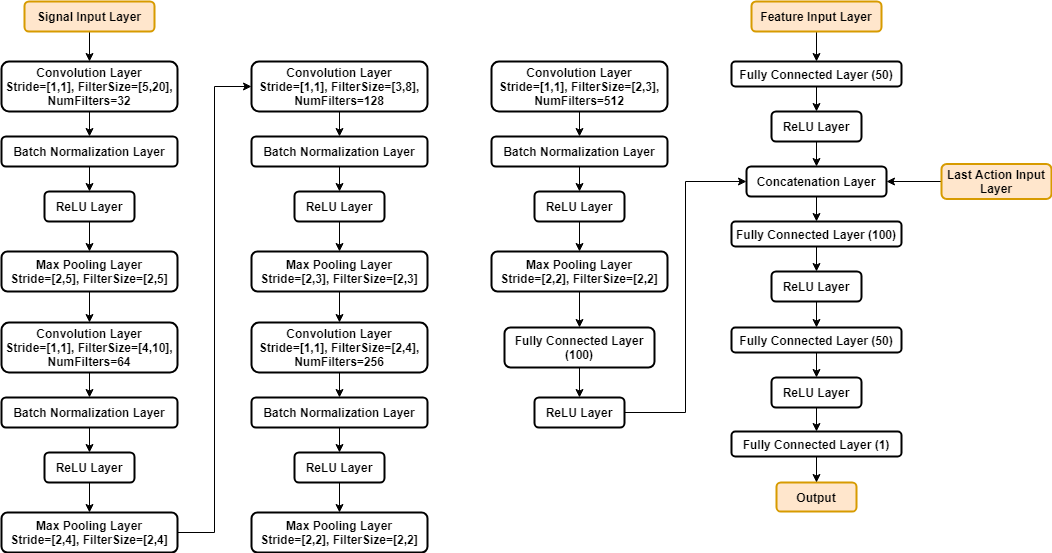}%
  \label{CriticNet}}
  \caption{The architecture of two networks using in weight-guided agent. (a) Actor network. (b) Critic network.}
  \label{RL_Net}
  \end{figure*}

The intelligent system comprises two operational components and two evaluative units. 
The actors, denoted as $T_1$ and $T_2$, calculate the probability distribution of output loss elements consonant with the input state, 
operating distinctively from the critics, which are constituted by separate neural networks designated by $V_1$ and $V_2$. 
The actor network $T_1$ computes the motion probability distribution requisite for the prevailing state, 
whereas actor $T_2$ is tasked with ascertaining the projected action distribution for the ensuing state. 
The critic network $V_1$ appraises the performance metric for the current action, 
and critic network $V_2$ evaluates the anticipated action score at the subsequent juncture. 
The architectural configuration of the actor is portrayed in Fig.\ref{ActorNet}, 
whereas the structural composition of the critic is exhibited in Fig.\ref{CriticNet}.

\subsubsection{Reward mechanism}

\begin{figure}[!htb]
  \centering
  \includegraphics[width=3.5in]{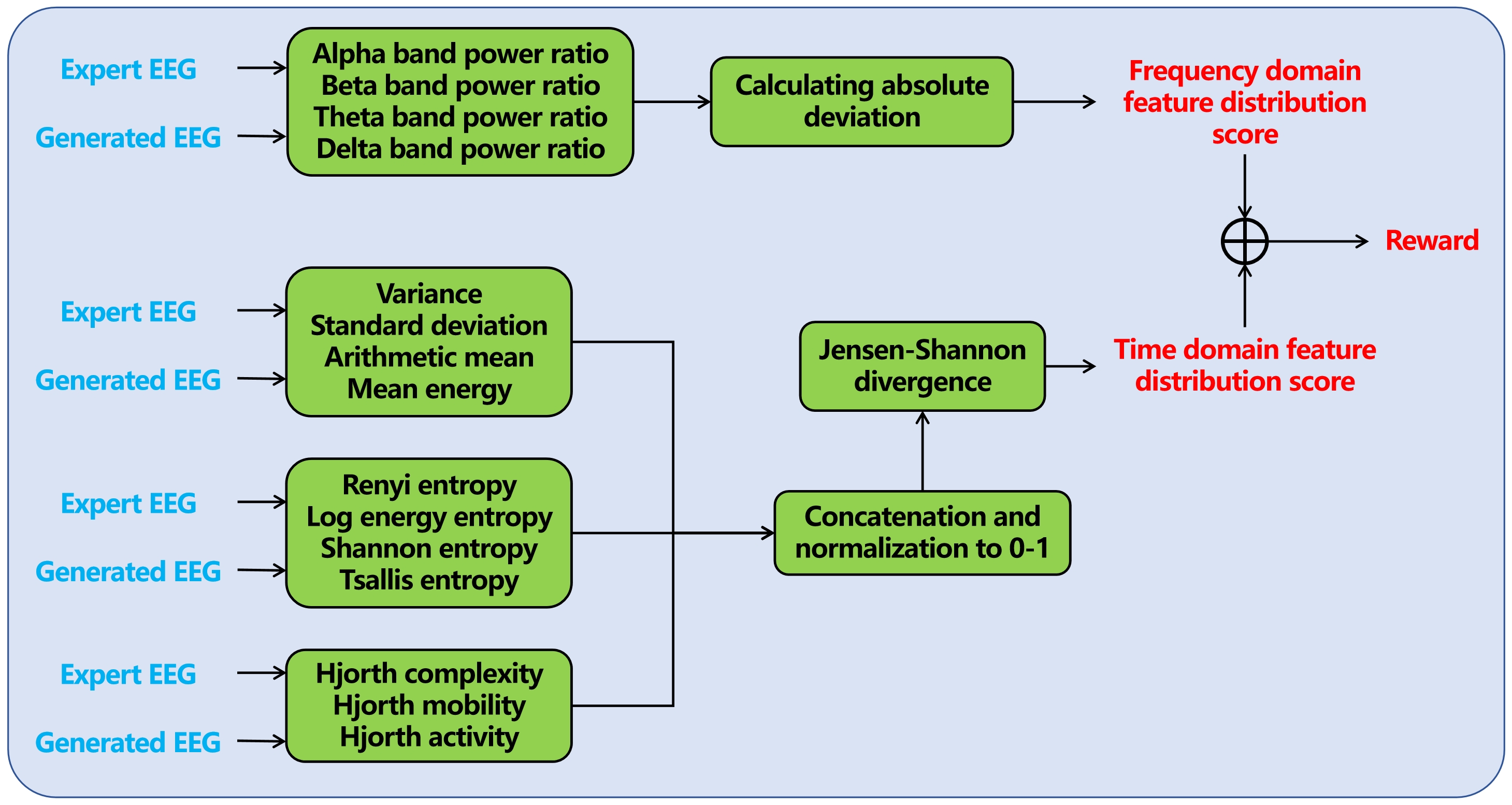}
  \caption{The design process of a reward mechanism for the EEG generation system.}
  \label{reward}
\end{figure}

The execution of an efficacious reward mechanism is imperative for augmenting learning efficacy. A meticulously designed reward function precisely articulates the objectives of the task, thereby enabling the intelligent system to acquire mechanisms for the efficient completion of tasks. The aim of the weight-guided agent is to ascertain the optimal weights for updating the EEG diffusion model, ensuring that the engendered brain electrical signals are in concordance with both temporal and spectral attributes. Consequently, the reward is bifurcated into two components. Initially, compute the scores for disparities in the temporal feature distribution. Subsequently, ascertain the score for divergences in the spectral feature distribution. We establish a standard electrical signal as the benchmark, and thereafter calculate the time-frequency deviation between the generated signal and the reference signal, as illustrated in Fig.\ref{reward}.

The temporal attributes of cerebral signals are categorized into three distinct groups: statistical, informational, and nonlinear characteristics. We utilize statistical parameters derived from brain signals, including variance, standard deviation, arithmetic mean, and mean energy, to quantify the statistical dimensions. Furthermore, we employ measures such as Renyi entropy, log energy entropy, Shannon entropy, and Tsallis entropy to delineate the informational content of the signal. To assess the nonlinear characteristics, we deploy metrics such as Hjorth complexity, Hjorth mobility, and Hjorth activity. To ascertain the divergence between probability distributions, we apply the Jensen-Shannon Divergence (JS divergence). The JS divergence is scaled from 0 to 1, with a score of 0 signifying identical distributions and a score of 1 indicating substantial disparity. In consideration of these factors, we amalgamate these features and calculate the JS divergence for each category, employing the scattering value as the metric that quantifies the distribution of temporal features.

The spectral characteristics of EEG signals involve the transformation of EEG signal attributes from their native temporal domain into spectral domain feature information. These can be categorized into five fundamental rhythms: alpha, beta, theta, delta, and gamma. Each frequency band within these classifications exhibits unique properties. Notably, gamma waves exert minimal influence in tasks pertaining to motor imagery. Accordingly, we compute the power spectrum for the initial four frequency bands and transform their energy into the proportion of energy within each frequency band. The absolute discrepancy between the original signal's energy proportion and that of the reference signal is utilized as the spectral domain feature distribution score. This score is then amalgamated with the temporal domain feature distribution score to constitute the ultimate reward.
\subsubsection{Update agent}

Initialize the critic $V(S,A,\phi)$ with random parameters $\phi$, and initialize the target critic $V_t(S,A,\phi_t)$ with the same parameters $\phi_t$. 
Initialize the actor $T(S,\psi)$ with random parameters $\psi$, and initialize the target actor $T_t(S,\psi_t)$ with the same parameters $\psi_t$. In each training step, the current observation is $S$, select the action:

\begin{equation}
  \label{Action}
  A = T(S,\phi) + N_s 
\end{equation}

where $N_s$ is the stochastic noise. Execute the action $A$. Observe the reward $R$ and the next observation $S'$. Store the experience $(S,A,R,S')$ in the experience buffer. Then, sample a random mini-batch of $B_m$ experiences $(S_i,A_i,R_i,S'_i)$ from the experience buffer. We can set the value function:

\begin{equation}
  \label{Value function}
  y_i = R_i + \tau V_t(S'_i,T_t(S'_i,\phi_t),\psi_t) 
\end{equation}

where $\tau$ represents the discount factor. 
The value function target comprises the aggregate of the immediate reward $R_i$ and the discounted subsequent reward. 
To ascertain the aggregate reward, the agent initially determines a subsequent action by conveying the ensuing observation $S'_i$ from the retrieved experience to the target actor. 
Prior to the revision of the actor and critic parameters, the preceding parameters of the actor and critic networks are replicated to their respective target networks. 
The agent then ascertains the cumulative reward by submitting the subsequent action to the target critic. The critic parameters are subsequently refined by minimizing the loss $L_c$ across the entire ensemble of sampled experiences.

\begin{equation}
  \label{Critic loss}
  L_c = -\frac{1}{2M}\sum_{n = 1}^{M}(y_i-V(S_i,A_i,\psi))^2    
\end{equation}

Update the actor parameters by maximizing the expected discounted reward. The loss function of the actor $L_a$ can be defined as:

\begin{equation}
  \label{Actor loss}
  L_a = -\frac{1}{M}\sum_{n = 1}^{M}V(S_i,T(S_i,\phi),\psi)^2 
\end{equation}

Collectively, the process of parameter updates throughout the actor-critic based EEG diffusion system is illustrated in algorithm \ref{algorithm}.
Initialize EEG-U-Net $E_Unet$ with parameters $\theta_u$. Initialize actor $T$ and target actor $T_t$ with parameters $\phi$ and $\phi_t$
Initialize critic $V$ and target critic $V_t$ with parameters $\psi$ and $\psi_t$. Set $t$ as the number of diffusion steps.
Set pre-trained wavelet network $W_{net}$ with parameters $\theta_w$.
Set pre-trained classification network $C_{net}$ with parameters $\theta_c$.
Set $L$ as the target label.
Set $\eta$ as the EEG-U-Net updating learning rate, $\alpha_a$ as the actor updating learning rate, 
$\alpha_c$ as the critic updating learning rate, $\upsilon$ as the discount factor

\begin{algorithm}[!htb]
  \caption{Update the models in EEG generation system}
  \label{algorithm}
  \begin{algorithmic}
  \STATE 
  \STATE \textbf{For each epoch}:
  \STATE \hspace{0.5cm} \textbf{For each training iteration}:
  \STATE \hspace{0.5cm} \hspace{0.5cm} $ \textbf{select randomly } x_0 \subset Dataset $
  \STATE \hspace{0.5cm} \hspace{0.5cm} $ x_t \gets ForwardDiffusion(x_0,t) $ Using Eq.(\ref{Forward diffusion process})
  \STATE \hspace{0.5cm} \hspace{0.5cm} $ \varepsilon_{pred} \gets E_{Unet}(x_t,L,t,\theta_u) $
  \STATE \hspace{0.5cm} \hspace{0.5cm} $ \textbf{generate } N_e \subset Gaussia Noise(0,1) $
  \STATE \hspace{0.5cm} \hspace{0.5cm} $ Y_1 \gets ReverseDiffusion(N_e,t,\theta_u) $ Using Eq.(\ref{Backward diffusion process})
  \STATE \hspace{0.5cm} \hspace{0.5cm} $ S \gets Merge(W_{net}(Y_1, \theta_w), C_{net}(Y_1, \theta_c)) $
  \STATE \hspace{0.5cm} \hspace{0.5cm} $ \textbf{generate } N_s \subset Gaussia Noise(0,1) $
  \STATE \hspace{0.5cm} \hspace{0.5cm} $ [w_d,w_{tf},w_c] \gets T(S,\phi) + N_s $
  \STATE \hspace{0.5cm} \hspace{0.5cm} $ Loss \gets w_d\times mse(x_0,\varepsilon_{pred})+w_{tf}\times crossentropy(W_{net}(Y_1,\theta_w),L)+ w_c\times crossentropy(C_{net}(Y_1, \theta_c),L) $
  \STATE \hspace{0.5cm} \hspace{0.5cm} $ \theta_u \gets \theta_u + \eta \cdot Gradient(Loss) $
  \STATE \hspace{0.5cm} \hspace{0.5cm} $ Y’_1 \gets ReverseDiffusion(N_e,t,\theta_u) $ Using Eq.(\ref{Backward diffusion process})
  \STATE \hspace{0.5cm} \hspace{0.5cm} $ S' \gets Merge(W_{net}(Y'_1, \theta_w), C_{net}(Y'_1, \theta_c)) $
  \STATE \hspace{0.5cm} \hspace{0.5cm} $ R \gets GetReward(S’,w_d,w_{tf},w_c) $
  \STATE \hspace{0.5cm} \hspace{0.5cm} $ \textbf{Store transition } (S, w_d,w_{tf},w_c, R, S') \textbf{ in buffer } B $

  \STATE \hspace{0.5cm} \textbf{End}
  \STATE \hspace{0.5cm}
  \STATE \hspace{0.5cm} $ \textbf{Sample a batch } (S_i,w_{d_i},w_{{tf}_i},w_{c_i},R_i,S'_i) \subset B $
  \STATE \hspace{0.5cm} $ TD_{error} \gets R_i + \tau \times V_t(S’_i,T_t(S'_i,\phi_t), \psi_t) - V(S_i,w_{d_i},w_{tf_i},w_{c_i},\psi) $
  \STATE \hspace{0.5cm} $ \phi \gets\phi + \alpha_a \times Gradient(-(V(S_i,T(S_i,\phi), \psi))^2) $
  \STATE \hspace{0.5cm} $ \psi \gets \psi+ \alpha_c \times Gradient(TD_{error}^2)$
  \STATE \hspace{0.5cm} $ \phi_t \gets \sigma \times \phi + (1 - \sigma) \times \phi_t $
  \STATE \hspace{0.5cm} $ \psi_t \gets \sigma \times \psi + (1 - \sigma) \times \psi_t $
  \STATE  \textbf{End}

  \end{algorithmic}
  \label{alg1}
\end{algorithm}

\section{Experiments and results}
\subsection{Dataset}
We have undertaken a comprehensive assessment of system performance, employing a synergistic approach that integrates both publicly accessible datasets, specifically the BCI Competition IV 2a dataset \cite{BCI4-2a}, and proprietary EEG data that we have amassed through our methodological frameworks. The specifics of this evaluation are delineated as follows:

{\bf BCI competition IV 2a:} The dataset under analysis consists of data from nine participants, each of whom performed four distinct motor imagery tasks, strictly following the prescribed protocols. The tasks included movements of the left hand, right hand, foot, and tongue. Each category was repeated 72 times, resulting in a total of 288 trials. Volunteers executed the imagined activities in accordance with the cues displayed on-screen. Preceding each task was a preparatory interval of two seconds, intended to prepare the participants for the impending motor imagery. This was followed by a four-second period allocated for the performance of the motor imagery. After each task, a pause of one and a half seconds was incorporated to allow for recovery. The data, collected at a sampling frequency of 250Hz, produced sequences of motor imagery data comprising 1000 sample points each. The dataset is composed of 25 channels, with 22 channels dedicated to EEG recordings, aligned with the international 10-20 system for electrode placement. The remaining three channels were allocated for electrooculography (EOG). For the purposes of our experimental analysis, our utilization was restricted to the 22 EEG channels, thereby ensuring that the focus of our investigation was on the cerebral electrical activity pertinent to the study.

{\bf In-house EEG Dataset (IHED):} In order to verify the effectiveness of the method proposed in this paper, we used RunE-HD high-density physiological signal acquisition device RT-R4132E to collect EEG signals of 5 subjects. 
The subject information is shown in the Table \ref{SubjectInfo}. 

\begin{table}[!htb]
  \caption{Details of recruited subjects}
  \label{SubjectInfo}
  \centering
  \setlength{\tabcolsep}{7mm}{}
  \begin{tabular}{|c||c|c|c|}
    \hline
    Subject ID & Age & Gender\\
    \hline
    S1 & 30 & Male\\
    \hline
    S2 & 26 & Male\\
    \hline
    S3 & 29 & Female\\
    \hline
    S4 & 23 & Male\\
    \hline
    S5 & 28 & Female\\
    \hline
  \end{tabular}
\end{table}

\begin{figure}[!htb]
  \centering
  \includegraphics[width=2in]{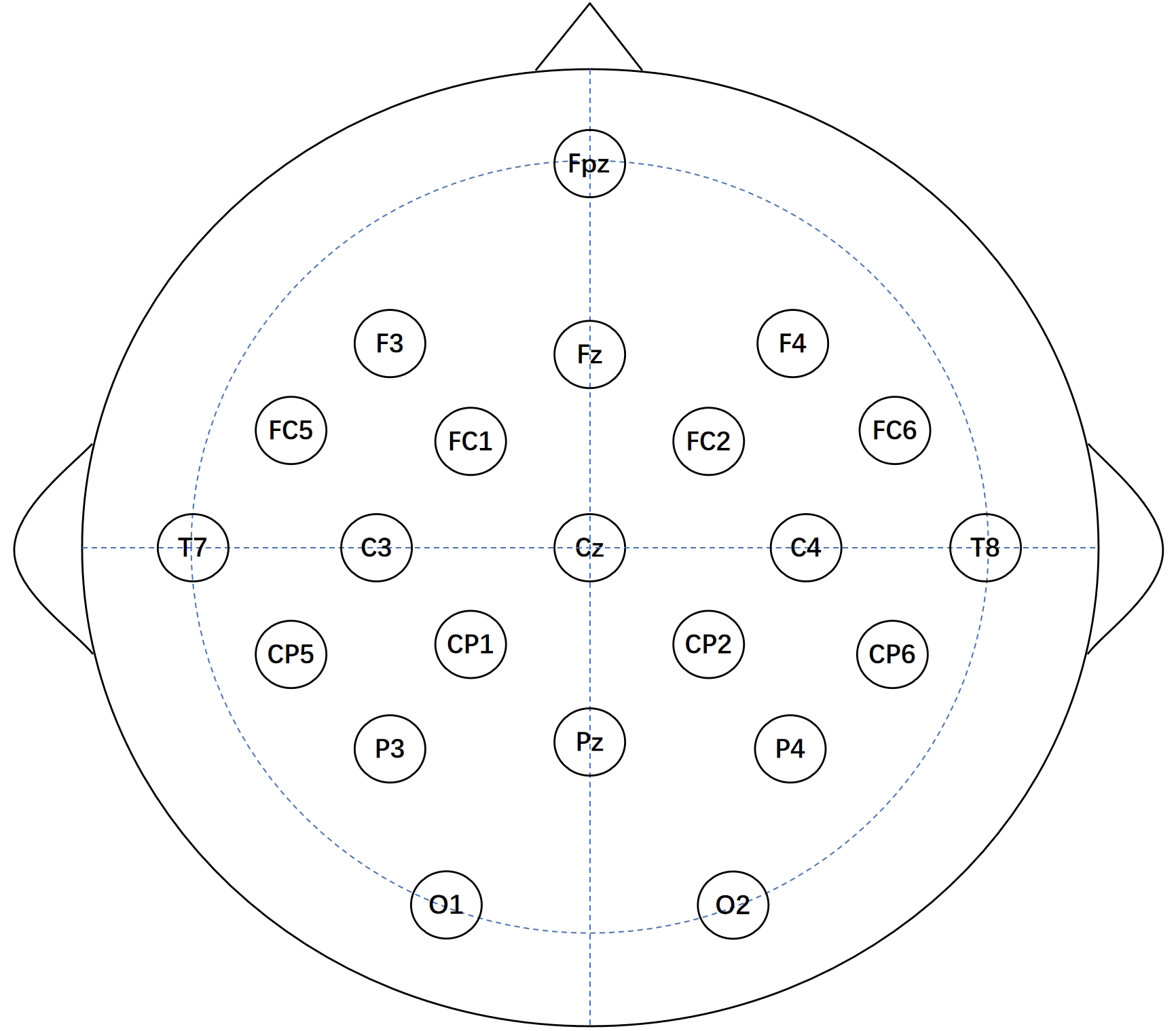}
  \caption{Electrode montage for In-house EEG Datase.}
  \label{Channel}
\end{figure}

\begin{figure}[!htb]
  \centering
  \includegraphics[width=3.5in]{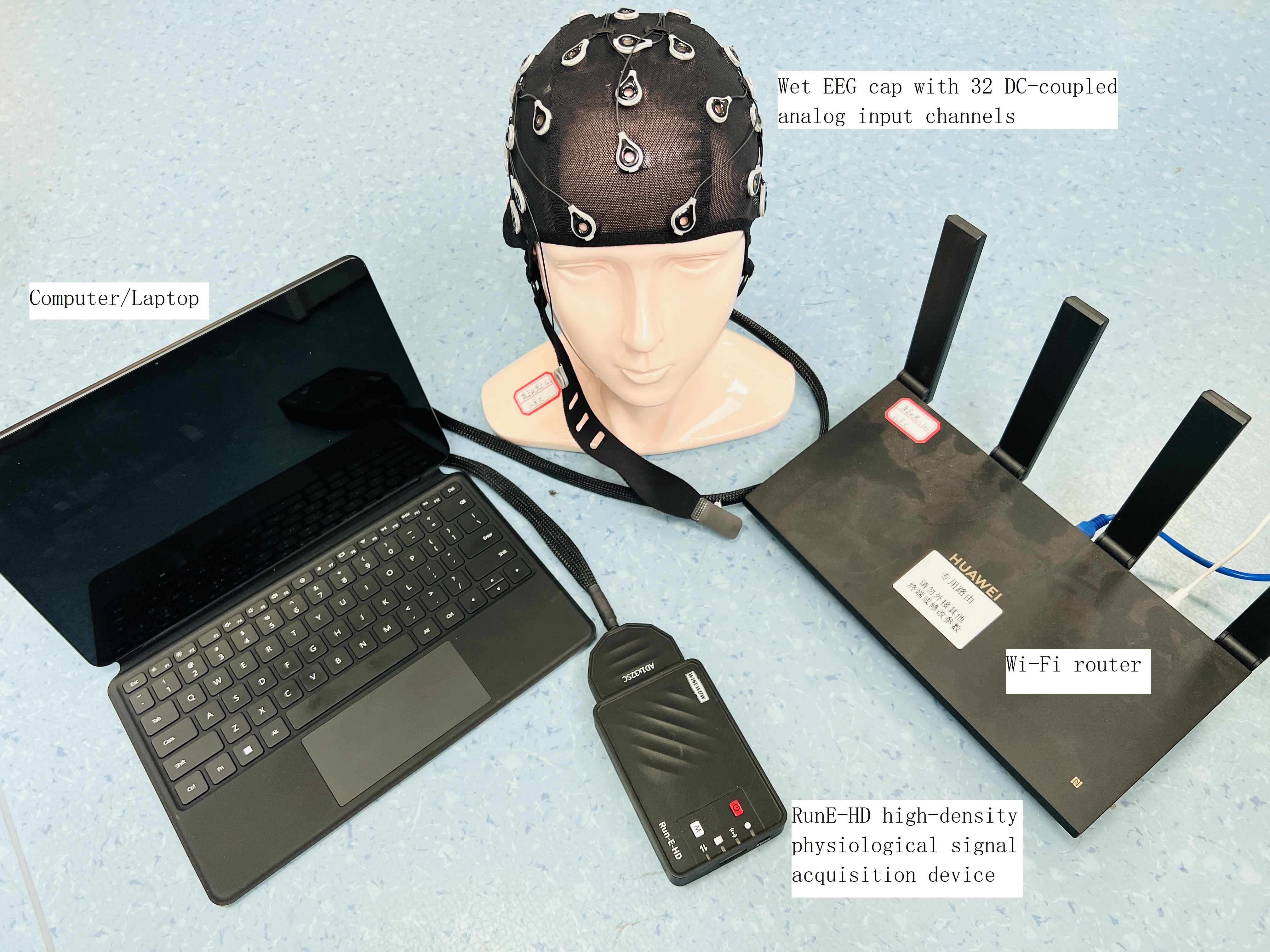}
  \caption{Equipments utilized for data collection.}
  \label{Device}
\end{figure}

The experiment was approved by Shandong First Medical University under document number “R202402280064”. 
It is a wet device that contains 32 DC-coupled analog input channels with 24 Bit resolution. 
We selected 22 relevant channels for the experiment, and the channel distribution is shown in the Fig.\ref{Channel}. 
The sampling frequency of the device is 500Hz, and we downsample the data to 250Hz.
After amplification, filtering, and AD conversion, the signals collected by the device are transmitted to the PC software via WIFI. 
The devices are shown in the Fig.\ref{Device}. 

\begin{figure}[!htb]
  \centering
  \includegraphics[width=3.5in]{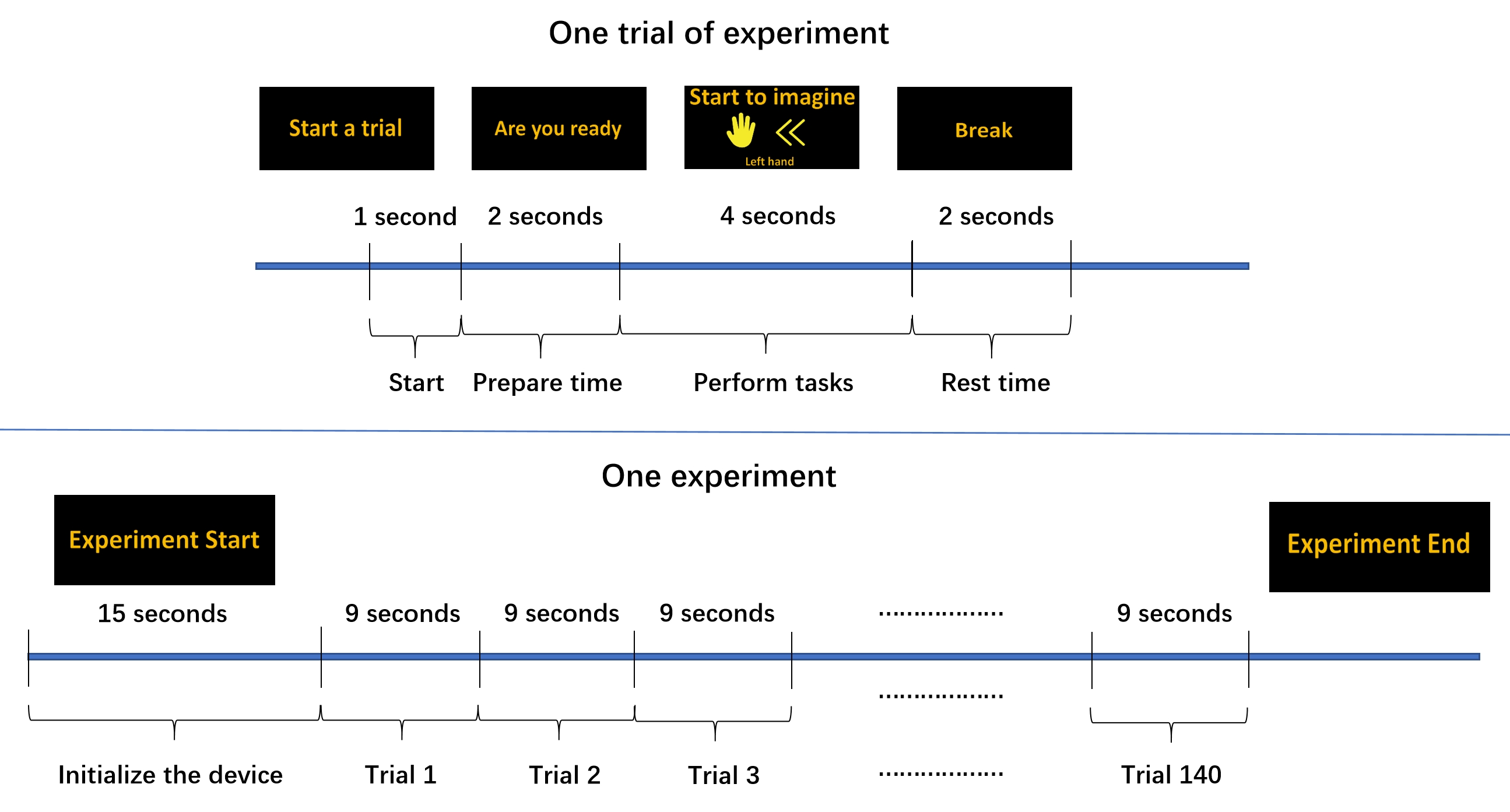}
  \caption{Timing scheme of one session (top), Timing scheme of the paradigm (bottom).}
  \label{Experiment}
\end{figure}

Participants completed the designated motor imagery tasks in accordance with the on-screen instructions. 
The experimental design for each set of trials is depicted in Fig.\ref{Experiment}. Commencing each set of trials, the initial 15 seconds marked the onset of the experiment, during which the screen displayed the message "Experiment Start." Subsequently, participants were instructed to randomly execute 70 imagined movements with their left hand and 70 with their right hand. Each data collection interval was set at 9 seconds, preceded by a 1-second preparatory prompt. At the experiment's commencement, participants were prompted to prepare for 2 seconds. Thereafter, the screen presented specific imaginary tasks at random. Participants were allotted 4 seconds to perform each task. Following this, a 2-second rest period was observed. This sequence was reiterated, resulting in the collection of 140 motor imagery data points per participant per set of trials.

\subsection{Preprocessing}
Before the commencement of the experimental procedure, all EEG signals are subjected to a rigorous preprocessing regimen. This process includes frequency domain filtering, removal of noise and artifacts, and normalization of the signals. The steps involved are as follows:
Initially, a bandpass filter is applied to refine the raw EEG signal within the frequency spectrum of 0.5 to 40 Hz. The principal aim of this step is to remove high-frequency noise and to eliminate any persistent direct current components.
Subsequently, considering the potential influence of eye and head movements on signal integrity, independent component analysis (ICA) is utilized to remove these artifacts and to mitigate the interference caused by motion.
Lastly, to correct for any discrepancies arising from electrode contact with the scalp, the re-referencing technique is implemented for signal calibration.

\subsection{Experiment A}
The primary objective of the experiment was to ascertain whether the EEG data produced by the proposed actor-critic based EEG diffusion system could significantly enhance the classification accuracy. For this purpose, we employed the auto-selected filter bank regularized common spatial pattern (ACSP) as the feature extraction technique and a single-channel series convolutional neural network (SCS-CNN) as the classifier \cite{DAFBCSP}. The dataset was meticulously divided into separate training and testing subsets.To establish a comparative baseline, the classifier was initially trained on the training dataset and then assessed on the test dataset.

In the initial experiment, we amalgamated the data from all subjects within each dataset to train the actor-critic based EEG diffusion system. The models in this system was trained on the same training data, with a batch size of 100 and 1000 iterations. We randomly generated 1000 EEG samples using this system and incorporated them into the original training data for feature extraction and model training. This enriched dataset was subsequently tested on the same test set. Furthermore, we performed a comparative analysis of our results with those obtained from other existing methods. The experimental outcomes and comparative results are presented in Table \ref{Experiment A}.

\begin{table*}[!htb]
  \caption{The experimental outcomes and comparative results for experiment A}
  \label{Experiment A}
  \centering
  \setlength{\tabcolsep}{7mm}{}
  \begin{tabular}{|c||c|c|c|c|c|c|}
    \hline
     & \multicolumn{3}{|c|}{BCI competition IV 2a} & \multicolumn{3}{|c|}{IHED} \\
    \hline
      & Accuracy & Kappa & F1-Score & Accuracy & Kappa & F1-Score\\
    \hline
    CSP \cite{CSP}                 & 63.27\% & 0.5103 & 0.6324 & 83.57\% & 0.6714 & 0.8356\\
    \hline
    ResNet \cite{ResNet}           & 67.82\% & 0.5710 & 0.6784 & 87.57\% & 0.7514 & 0.8752\\
    \hline
    CP-MixedNet \cite{CP-MixedNet} & 70.06\% & 0.6008 & 0.6997 & 88.43\% & 0.7686 & 0.8841\\
    \hline
    3DCNN \cite{3DCNN}             & 72.15\% & 0.6286 & 0.7213 & 87.57\% & 0.7514 & 0.8757\\
    \hline
    EEGNet \cite{EEGNet}           & 74.02\% & 0.6535 & 0.7413 & 89.86\% & 0.7971 & 0.8985\\
    \hline
    Deep ConvNet \cite{Shallow}    & 77.85\% & 0.7047 & 0.7785 & 91.29\% & 0.8257 & 0.9128\\
    \hline
    Shallow ConvNet \cite{Shallow} & 80.26\% & 0.7368 & 0.8023 & 93.00\% & 0.8600 & 0.9300\\
    \hline
    ACSP+SCS-CNN \cite{DAFBCSP}        & 81.10\% & 0.7479 & 0.8109 & 94.29\% & 0.8857 & 0.9429\\
    \hline
    Proposed method                & 84.41\% & 0.7922 & 0.8442 & 95.43\% & 0.9086 & 0.9543\\
    \hline
  \end{tabular}
\end{table*}

\begin{table*}[!htb]
  \caption{The experimental outcomes and comparative results for experiment B\label{Experiment B}}
  \centering
  \setlength{\tabcolsep}{7mm}{}
  \begin{tabular}{|c||c|c|c|c|c|c|}
    \hline
     & \multicolumn{3}{|c|}{BCI competition IV 2a} & \multicolumn{3}{|c|}{IHED} \\
    \hline
      & Accuracy & Kappa & F1-Score & Accuracy & Kappa & F1-Score\\
    \hline
    CSP \cite{CSP}                  & 59.14\% & 0.4552 & 0.5906 & 81.57\% & 0.6314 & 0.8157\\
    \hline
    ResNet \cite{ResNet}            & 64.20\% & 0.5226 & 0.6414 & 85.86\% & 0.7171 & 0.8586\\
    \hline
    CP-MixedNet \cite{CP-MixedNet}  & 66.74\% & 0.5566 & 0.6672 & 87.71\% & 0.7543 & 0.8771\\
    \hline
    3DCNN \cite{3DCNN}              & 69.02\% & 0.5869 & 0.6896 & 86.29\% & 0.7257 & 0.8628\\
    \hline
    EEGNet \cite{EEGNet}            & 71.99\% & 0.6265 & 0.7199 & 89.57\% & 0.7914 & 0.8957\\
    \hline
    Deep ConvNet \cite{Shallow}     & 75.66\% & 0.6754 & 0.7568 & 89.43\% & 0.7886 & 0.8941\\
    \hline
    Shallow ConvNet \cite{Shallow}  & 77.47\% & 0.6996 & 0.7744 & 91.86\% & 0.8371 & 0.9185\\
    \hline
    ACSP+SCS-CNN \cite{DAFBCSP}         & 79.32\% & 0.7243 & 0.7926 & 92.14\% & 0.8429 & 0.9213\\
    \hline
    Proposed method                 & 82.10\% & 0.7613 & 0.8211 & 93.43\% & 0.8686 & 0.9343\\
    \hline
  \end{tabular}
\end{table*}

\subsection{Experiment B}

The results indicate that the classification performance of the model, when trained with data generated by the hybrid EEG generation system, is markedly superior to that of the model trained with the original data. Relative to other existing methods, the approach detailed in this paper produces the most optimal outcomes. These findings validate that the data augmentation technique presented herein effectively improves the classifier's performance.

In the subsequent experiment, we continue to employ the ACSP as the feature extraction method and SCS-CNN as the classifier. The distinction in this instance is the utilization of data from each single participant as the test set, while data from the remaining participants constitute the training set. This methodology is designed to demonstrate that the integration of generated data can enhance the generalization capability of the classification model, thereby indirectly validating the comprehensiveness of the data produced by the actor-critic based EEG diffusion system.
We train the actor-critic based EEG diffusion system using the aforementioned training data, with a batch size of 100 and 1000 training iterations. The system then randomly generates 1000 data samples, which are integrated with the original training data for concurrent feature extraction and classifier training. This combined dataset is subsequently evaluated on the test set.
Furthermore, we also conducted a comparative analysis of our approach with other existing methods. The experimental results and comparative analysis are detailed in Table \ref{Experiment B}.

According to the findings of our analysis, it is apparent that the synthesized data substantially enriches the diversity of the dataset, enabling the classification model to discern more prevalent features inherent in EEG signals. Consequently, this enhancement augments the model's generalization capabilities. This improvement enables the model to demonstrate greater proficiency in recognizing novel data. When compared with contemporary methodologies, the approach presented in this scholarly work has achieved the most favorable results, thereby validating the efficacy and superiority of the proposed method over existing techniques.

\section{Discussion}

\subsection{Quantitative analysis}

\begin{table*}[!htb]
  \caption{Analysis of the proposed method for evaluating the outcomes of individual components with the aim of enhancing the performance of the classification model (BCI competition IV 2a dataset).}
  \label{Quantitative analysis BCI4-2a}
  \centering
  \setlength{\tabcolsep}{3.5mm}{}
  \begin{tabular}{|c||c|c|c|c|c|c|c|c|c|}
    \hline
     & \multicolumn{3}{|c|}{Only use original data} & \multicolumn{3}{|c|}{\makecell{Original data + \\ EEG diffusion model}} & \multicolumn{3}{|c|}{\makecell{Original data +\\ EEG diffusion model +\\ Weight-guided agent}} \\
    \hline
      & Accuracy & Kappa & F1-Score & Accuracy & Kappa & F1-Score & Accuracy & Kappa & F1-Score\\
    \hline
    Sub 1         & 88.54\% & 0.8472 & 0.8854 & 89.24\% & 0.8565 & 0.8925 & 89.58\% & 0.8611 & 0.8957\\
    \hline
    Sub 2         & 72.66\% & 0.6355 & 0.7274 & 76.39\% & 0.6852 & 0.7639 & 78.82\% & 0.7176 & 0.7881\\
    \hline
    Sub 3         & 95.14\% & 0.9352 & 0.9515 & 95.49\% & 0.9398 & 0.9550 & 96.53\% & 0.9537 & 0.9653\\
    \hline
    Sub 4         & 58.33\% & 0.4444 & 0.5836 & 63.54\% & 0.5139 & 0.6353 & 63.19\% & 0.5093 & 0.6317\\
    \hline
    Sub 5         & 65.62\% & 0.5417 & 0.6559 & 68.06\% & 0.5741 & 0.6806 & 68.40\% & 0.5787 & 0.6837\\
    \hline
    Sub 6         & 54.51\% & 0.3935 & 0.5451 & 64.58\% & 0.5278 & 0.6465 & 66.32\% & 0.5509 & 0.6632\\
    \hline
    Sub 7         & 94.44\% & 0.9259 & 0.9443 & 94.79\% & 0.9306 & 0.9479 & 94.79\% & 0.9306 & 0.9481\\
    \hline
    Sub 8         & 90.28\% & 0.8704 & 0.9029 & 90.62\% & 0.8750 & 0.9064 & 90.97\% & 0.8796 & 0.9103\\
    \hline
    Sub 9         & 89.24\% & 0.8565 & 0.8923 & 90.28\% & 0.8704 & 0.9026 & 92.01\% & 0.8935 & 0.9202\\
    \hline
    \makecell{Mean\\$\pm$std}  & \makecell{78.75\%\\$\pm$16.07\%} & \makecell{0.7167\\$\pm$0.2143} & \makecell{0.7876\\$\pm$0.1607} & \makecell{81.44\%\\$\pm$13.27\%} & \makecell{0.7526\\$\pm$0.1769} & \makecell{0.8145\\$\pm$0.1326} & \makecell{82.29\%\\$\pm$13.26\%} & \makecell{0.7639\\$\pm$0.1768} & \makecell{0.8229\\$\pm$0.1328}\\
    \hline
    P-value       & /\ & /\ & /\ & 0.0383 & 0.0384 & 0.0388 & 0.0199 & 0.0199 & 0.0197\\
    \hline
  \end{tabular}
\end{table*}

\begin{table*}[!htb]
  \caption{Analysis of the proposed method for evaluating the outcomes of individual components with the aim of enhancing the performance of the classification model (In-house EEG dataset).}
  \label{Quantitative analysis IHED}
  \centering
  \setlength{\tabcolsep}{3.5mm}{}
  \begin{tabular}{|c||c|c|c|c|c|c|c|c|c|}
    \hline
     & \multicolumn{3}{|c|}{Only use original data} & \multicolumn{3}{|c|}{\makecell{Original data + \\ EEG diffusion model}} & \multicolumn{3}{|c|}{\makecell{Original data +\\ EEG diffusion model +\\ Weight-guided agent}} \\
    \hline
      & Accuracy & Kappa & F1-Score & Accuracy & Kappa & F1-Score & Accuracy & Kappa & F1-Score\\
    \hline
    S1           & 95.50\% & 0.9100 & 0.9550 & 96.29\% & 0.9257 & 0.9628 & 96.86\% & 0.9371 & 0.9686\\
    \hline
    S2           & 93.07\% & 0.8614 & 0.9307 & 94.36\% & 0.8871 & 0.9436 & 95.29\% & 0.9057 & 0.9529\\
    \hline
    S3           & 87.50\% & 0.7500 & 0.8749 & 89.79\% & 0.7957 & 0.8978 & 90.14\% & 0.8029 & 0.9014\\
    \hline
    S4           & 91.57\% & 0.8314 & 0.9157 & 92.57\% & 0.8514 & 0.9257 & 93.14\% & 0.8629 & 0.9314\\
    \hline
    S5           & 86.86\% & 0.7371 & 0.8686 & 89.86\% & 0.7971 & 0.8985 & 89.93\% & 0.7986 & 0.8993\\
    \hline
    \makecell{Mean\\$\pm$std} & \makecell{90.90\%\\$\pm$3.68\%} & \makecell{0.8180\\$\pm$0.0736} & \makecell{0.9090\\$\pm$0.0368} & \makecell{92.57\%\\$\pm$2.83\%} & \makecell{0.8514\\$\pm$0.0567} & \makecell{0.9257\\$\pm$0.0284} & \makecell{93.07\%\\$\pm$3.07\%} & \makecell{0.8614\\$\pm$0.0614} & \makecell{0.9307\\$\pm$0.0307}\\
    \hline
    P-value      & /\ & /\ & /\ & 0.0163 & 0.0164 & 0.0163 & 0.0025 & 0.0025 & 0.0025\\
    \hline
  \end{tabular}
\end{table*}

The methodology we propose is primarily composed of two integral components: the diffusion model and the reinforcement learning model. The diffusion model is chiefly responsible for generating the target datasets, while the reinforcement learning model plays a pivotal role in guiding the generator's update strategy. It refines the quality of the data produced by the generator by adjusting the network parameters. In the subsequent section, we will explore the significance of the reinforcement learning model within the framework of our proposed approach.

We employ the ten-fold cross-validation method to randomly partition the original data into separate training and testing sets. The classifier training process adheres to three distinct protocols: Firstly, feature extraction and classifier training are conducted exclusively with the original training data, followed by the assessment of ten classification accuracies on each test set of each subject. These outcomes serve as the reference benchmarks. Secondly, the squared error is utilized as the loss function for the training of the diffusion model. An additional 400 data samples are randomly generated and integrated with the original training dataset, facilitating a concurrent feature extraction and classifier training procedure. In the third approach, a composite loss function is employed, incorporating squared error, classification loss, and time-frequency loss. A reinforcement learning model is incorporated to dynamically adjust the weights during the training of the diffusion model. Another set of 400 data samples is randomly generated and merged with the original training data for a collective feature extraction and classifier training endeavor. Upon comparing these three methodologies and subjecting them to cyclic testing with the same test data, 
the subsequent results are presented in Table \ref{Quantitative analysis BCI4-2a} and Table \ref{Quantitative analysis IHED}.

The results obtained indicate that the integration of a reinforcement learning model results in a significant improvement in the overall classification accuracy (p-value$<$0.05), suggesting that this approach enables the diffusion model to generate novel samples that closely resemble the original training samples. The newly developed model is more adept at understanding and incorporating the variability in data transformations during the training phase, thereby enhancing its ability to handle unseen data and improving the model's generalization capacity.
When a model focuses excessively on the existing, albeit limited, data points during the training process, it is susceptible to overfitting, essentially over-adjusting to the training data at the expense of its generalization ability. The introduction of additional variants and exposing the model to a wider range of inputs helps to counteract overfitting, thereby enhancing the model's ability to discern complex patterns, which in turn improves classification accuracy.

\subsection{Qualitative analysis}
There exists a significant divergence between the methodologies utilized for the generation of EEG signals and images. Image generation is predominantly focused on the nuances of pixel-level details and the visual coherence among images. In contrast, EEG signals are distinguished by time series data that exhibit intricate temporal relationships and physiological properties. As a result, while a generated signal may appear superficially similar to an EEG signal, it may not inherently possess a complete suite of EEG feature information. The generation of EEG signals requires consideration of the continuity and specific spectral characteristics inherent in physiological signals.

The proposed actor-critic based EEG diffusion system is developed to address this challenge. We have generated 1000 novel EEG signals, encompassing both left-hand and right-hand motor imagery categories, with each category consisting of 500 EEG signals. We have conducted a thorough analysis of these generated EEG signals, which includes time domain analysis, time-frequency analysis, and an examination of feature distributions.

\subsubsection{Time domain analysis}

\begin{figure}[!htb]
  \centering
  \includegraphics[width=3.5in]{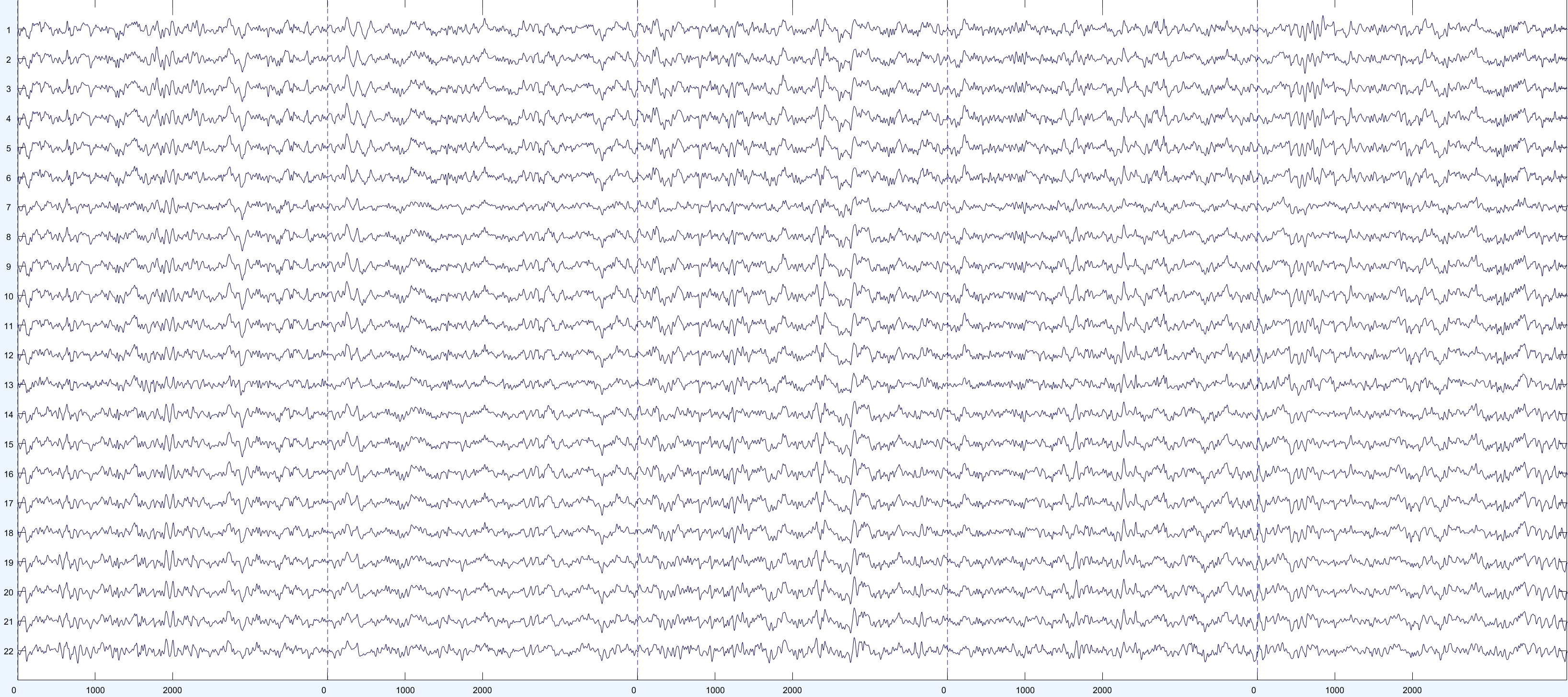}
  \caption{Real EEG signals drawn from BCI competition 4-2a dataset.}
  \label{RealEEG}
\end{figure}

\begin{figure}[!htb]
  \centering
  \includegraphics[width=3.5in]{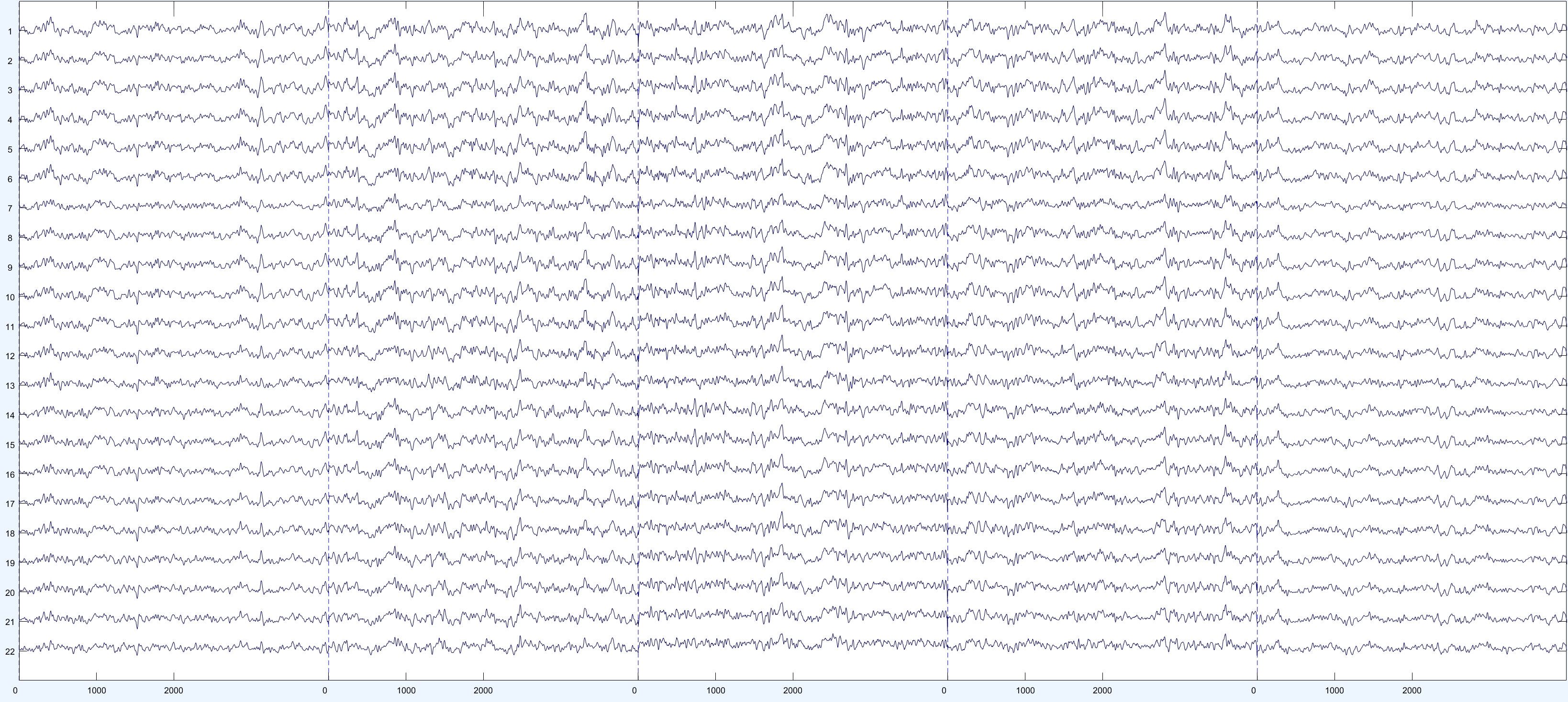}
  \caption{Generated EEG signals drawn from BCI competition 4-2a dataset.}
  \label{FakeEEG}
\end{figure}

The principal aim of EEG time domain analysis is to scrutinize the progression of waveform characteristics and oscillatory patterns within both synthetic and genuine signals. This analytical method provides a clear perspective on the fundamental properties of EEG dynamics and the anticipated responses to stimuli. As illustrated in Fig.\ref{RealEEG} and Fig.\ref{FakeEEG}, the time domain waveforms of various synthesized and authentic signals are depicted, offering a visual representation of these pivotal features.

Upon reviewing the results, it is evident that the waveform fluctuations of the generated signal closely mimic those of the authentic signal, with no discernible anomalies or deviations. The differentiation between the generated signal and the actual signal is not distinguishable based solely on waveform analysis. However, waveform analysis alone is not sufficient to determine whether the generated signal conforms to the intrinsic characteristics of EEG signals. Therefore, we apply an 8-13Hz bandpass filter to the signals associated with left-hand motor imagery and right-hand motor imagery. This specific frequency range is known to be movement-correlated. Within this band, we analyze the energy distribution of both signal types.

\begin{figure}[!htb]
  \centering
  \includegraphics[width=3.5in]{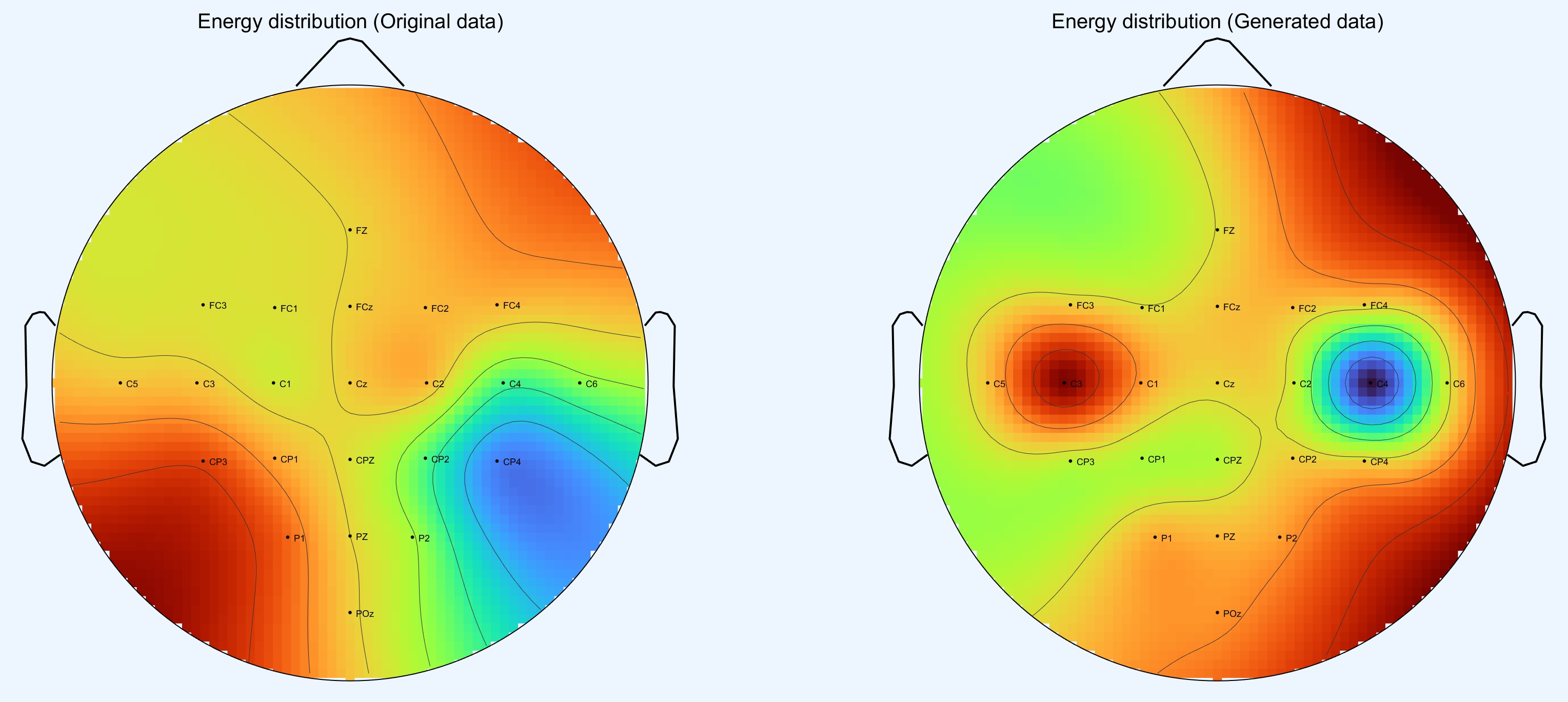}
  \caption{The energy distribution maps drawn from BCI competition 4-2a dataset. The distribution graph situated on the left has been constructed utilizing actual data, whereas the distribution graph situated on the right has been derived from synthetically generated data.}
  \label{Energy}
\end{figure}

We then calculate the variance for each channel of both the original and the generated signals, subsequently creating an EEG distribution map in accordance with the electrode layout. This map serves to illustrate the energy transfer process associated with motor imagery. Fig.\ref{Energy} presents the energy difference distribution maps for both the authentic and generated signals of imagined left-hand and right-hand movements, thereby highlighting the comparative energy distribution between the two.

Upon the subject's contemplation of left-hand movement, it is observed that the energy within the C3 channel is slightly elevated relative to that of the C4 channel. In contrast, when the subject envisions movement of the right hand, the energy of the C3 channel is found to be slightly lower than that of the C4 channel. The disparity distribution diagram indicates that the variance in energy for the authentic signals of motor imagery for both the left and right hands is predominantly concentrated within the C3 and C4 channels, demonstrating a reciprocal relationship where an increase in energy in one corresponds to a decrease in the other, in accordance with the expected conditions. Similar characteristics are discernible within the generated signal.
In conclusion, the generated signal exhibits the temporal domain attributes of the motor imagery signal, reflecting the anticipated energy dynamics between the C3 and C4 channels during the imagined motion of either hand.

\subsubsection{Frequency domain analysis}
\begin{figure}[!htb]
  \centering
  \includegraphics[width=3.5in]{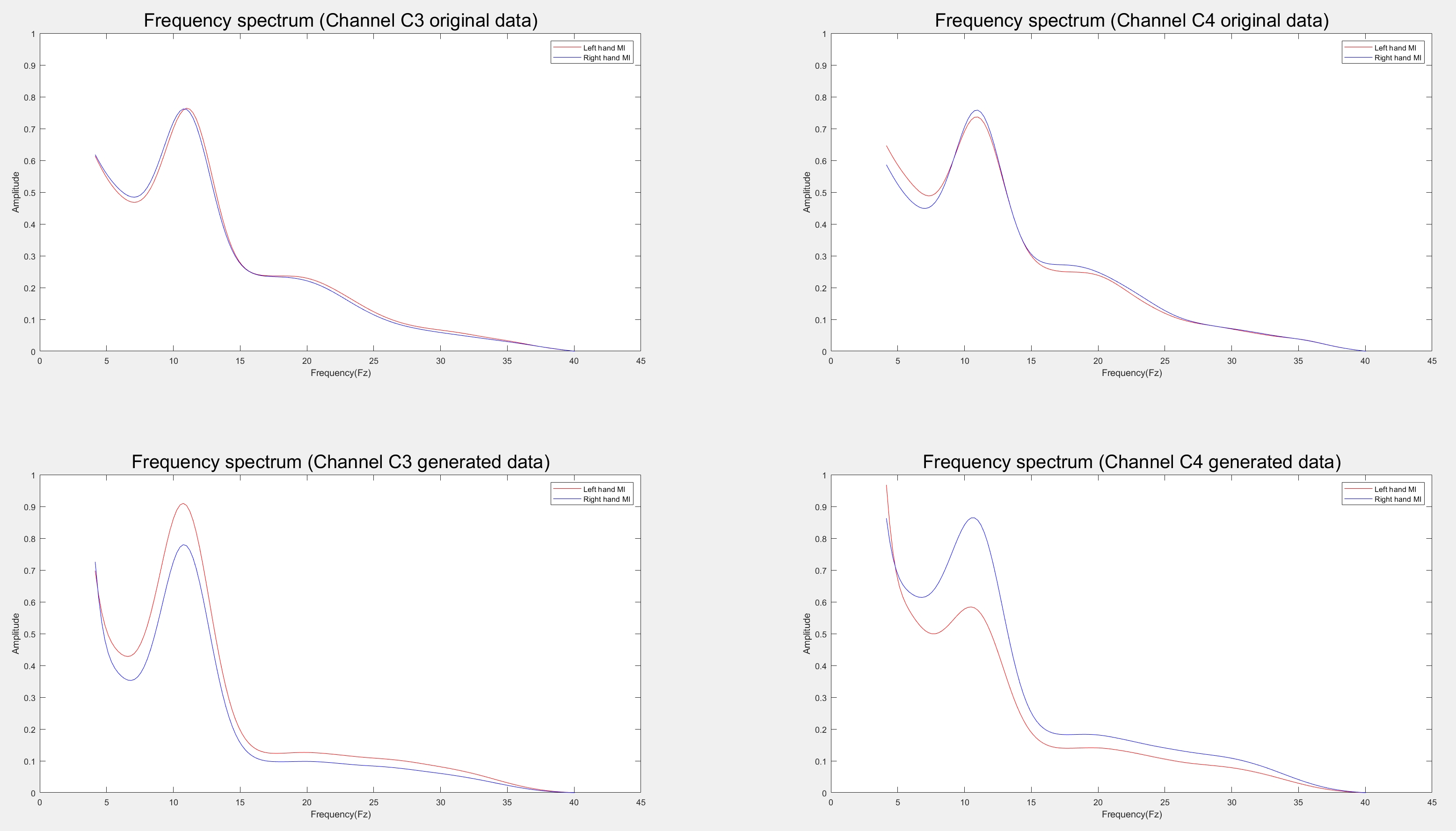}
  \caption{The spectral graph is derived from the data within BCI competition 4-2a dataset. The two spectrograms positioned at the upper section are derived from authentic data, and the two spectrograms at the lower section are similarly sourced from genuine data. The left-hand side depicts the spectral analysis of the motor imagery of the left hand, while the right-hand side illustrates the spectral analysis of the motor imagery of the right hand. The red and blue lines correspond to the C3 and C4 channels, respectively.}
  \label{Frequency}
\end{figure}

Frequency domain analysis enables the conversion of complex time domain signals into their corresponding frequency domain representations, thereby revealing the distinct frequency components inherent in the signal. This method is especially beneficial for identifying EEG activity patterns within a specific frequency band, which can be indicative of various brain states and functions. To determine the frequency domain energy associated with the left-hand motor imagery and the right-hand motor imagery signals, we utilized the Welch method. Thereafter, we computed the average of all the spectrograms to achieve a comprehensive representation.
As shown in Fig.\ref{Frequency}, this approach provides a clear visualization of the signal's spectrum.

Upon examination of the spectral graph, it is apparent that during the performance of left-hand motor imagery, the 8-13Hz frequency domain energy of the EEG signal recorded at the C4 channel shows a slight decrease compared to that at the C3 channel. In contrast, when the subject engages in right-hand motor imagery, the 8-13Hz frequency domain energy of the EEG signal at the C3 channel is observed to be marginally reduced in comparison to the C4 channel. These patterns are similarly observed in the characteristics of the generated signal.

\begin{figure}[!htb]
  \centering
  \includegraphics[width=3.5in]{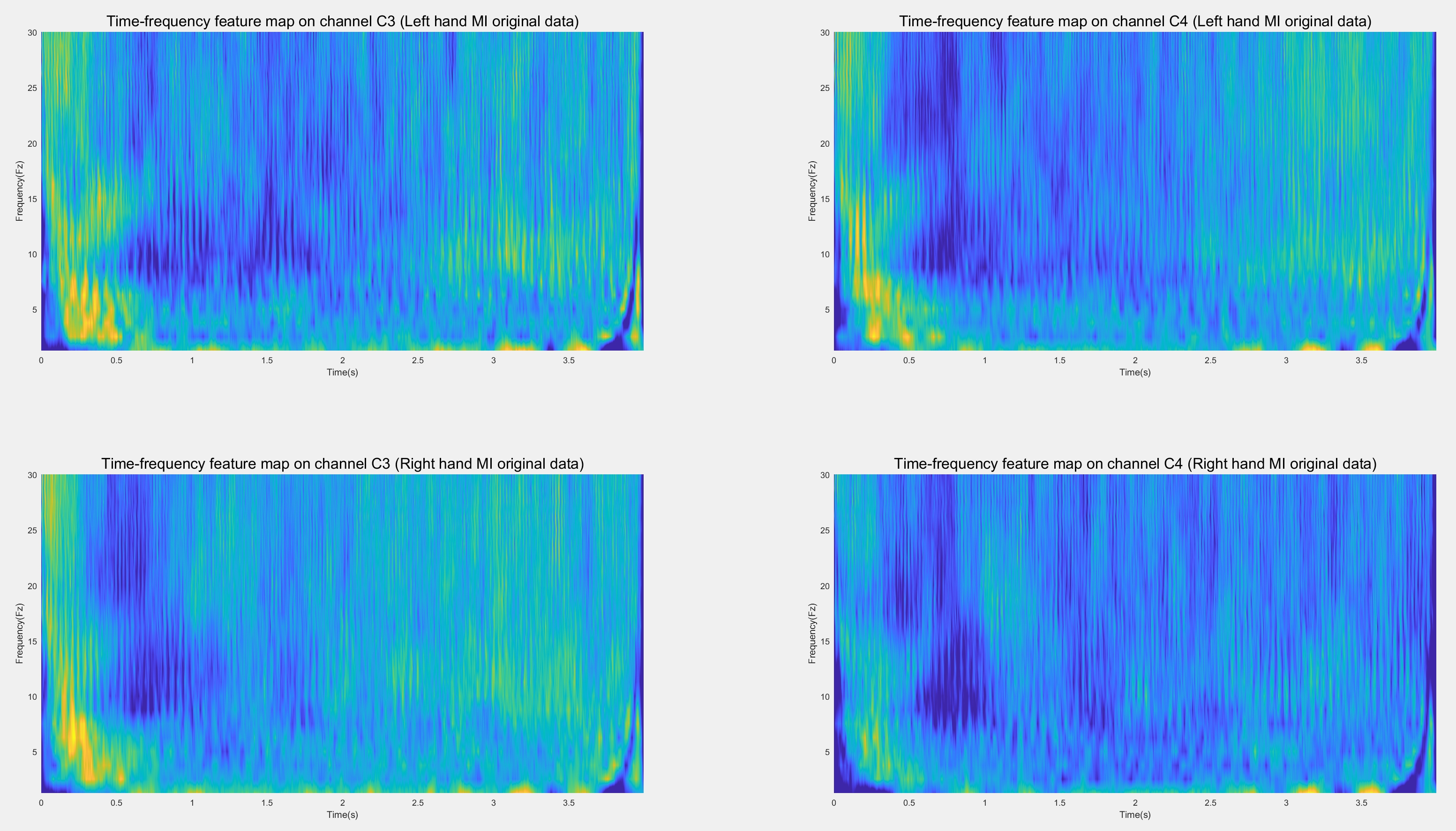}
  \caption{The time-frequency feature map is derived from the original data from BCI competition 4-2a dataset. The two upper figures represent the time-frequency feature maps for the C3 and C4 channels corresponding to the motor imagery of the left hand. The two lower figures depict the time-frequency feature maps for the C3 and C4 channels corresponding to the motor imagery of the right hand.}
  \label{Time_Frequency1}
\end{figure}

\begin{figure}[!htb]
  \centering
  \includegraphics[width=3.5in]{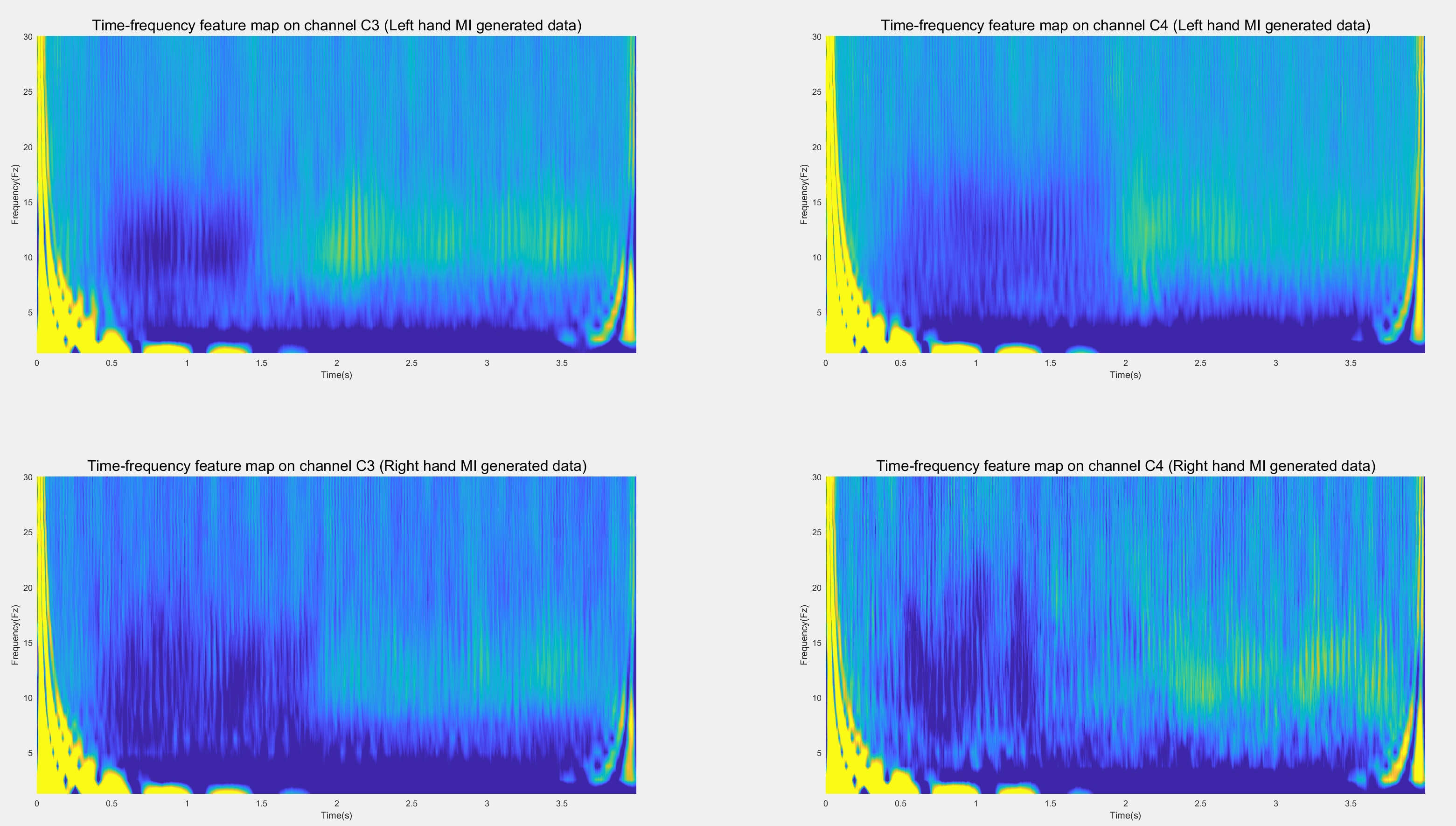}
  \caption{The time-frequency feature maps are computed from the synthetic data generated for BCI competition 4-2a dataset. The two upper illustrations portray the time-frequency feature maps for the C3 and C4 channels associated with the motor imagery of the left hand. The two lower illustrations depict the time-frequency feature maps for the C3 and C4 channels associated with the motor imagery of the right hand.}
  \label{Time_Frequency2}
\end{figure}

Time-frequency analysis combines the dual dimensions of time and frequency to dissect brain activity, thereby enabling a comprehensive examination of the temporal dynamics and frequency components within EEG signals. To extract time-frequency characteristics from both authentic and synthetic signals, we utilize the continuous wavelet transform.
This method generates a feature map, from which we compute the average value to derive the average time-frequency energy map, as depicted in Fig.\ref{Time_Frequency1} and Fig.\ref{Time_Frequency2}.

From the time-frequency energy plot, it is apparent that during the execution of the motor imagery task by the subjects, the C3 and C4 channels of the EEG signal demonstrate distinct Event-Related Desynchronization (ERD) and Event-Related Synchronization (ERS) phenomena within the low-frequency band. Similar features are also discernible in the generated signal, highlighting its conformity to the time-frequency domain characteristics typical of motor imagery signals.

\subsubsection{Feature distribution analysis}

The objective of generating motor imagery signals is to augment the training dataset, thereby enhancing the classifier's precision. Accordingly, we conduct an analysis of the feature distribution within the expanded dataset. We commence by utilizing EEG signal data to train a convolutional neural network for classifying movements of the left and right hands. Thereafter, we extract the final fully connected layer of the network, which encompasses the compressed features of the categorized data, and proceed to generate feature distribution maps for the motor imagery associated with both left and right hand movements. Additionally, we utilize the Fréchet Inception Distance (FID) to evaluate the divergence between the two distributions.

\begin{figure}[!htb]
  \centering
  \includegraphics[width=3.5in]{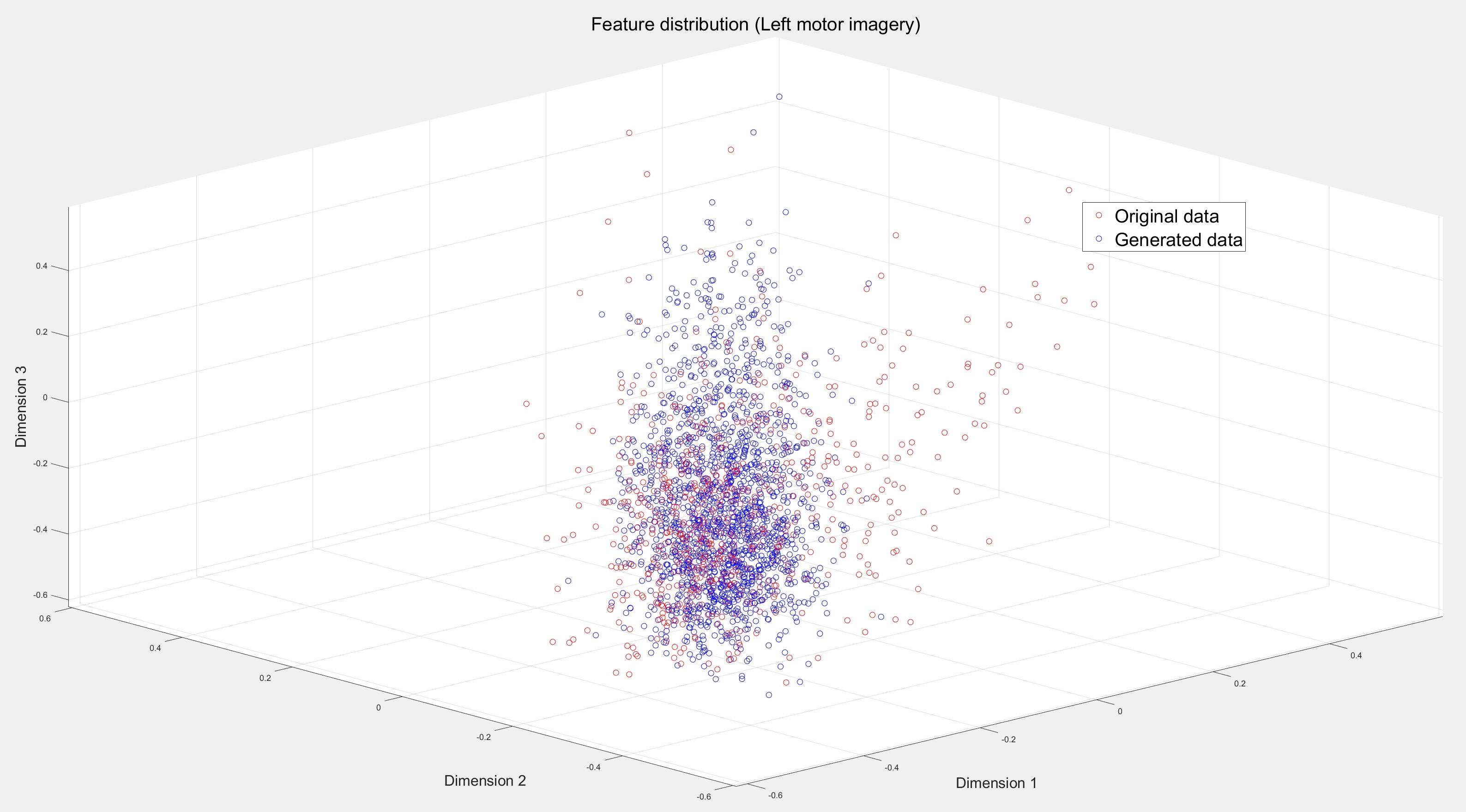}
  \caption{This illustration is a three-dimensional representation of the feature distribution for EEG signals (FID=0.0186), wherein the red data points denote the original data feature distribution associated with the motor imagery of the left hand from BCI competition 4-2a dataset, and the blue data points signify the feature distribution of the generated data corresponding to the motor imagery of the left hand from the same dataset.}
  \label{Distribution1}
\end{figure}

\begin{figure}[!htb]
  \centering
  \includegraphics[width=3.5in]{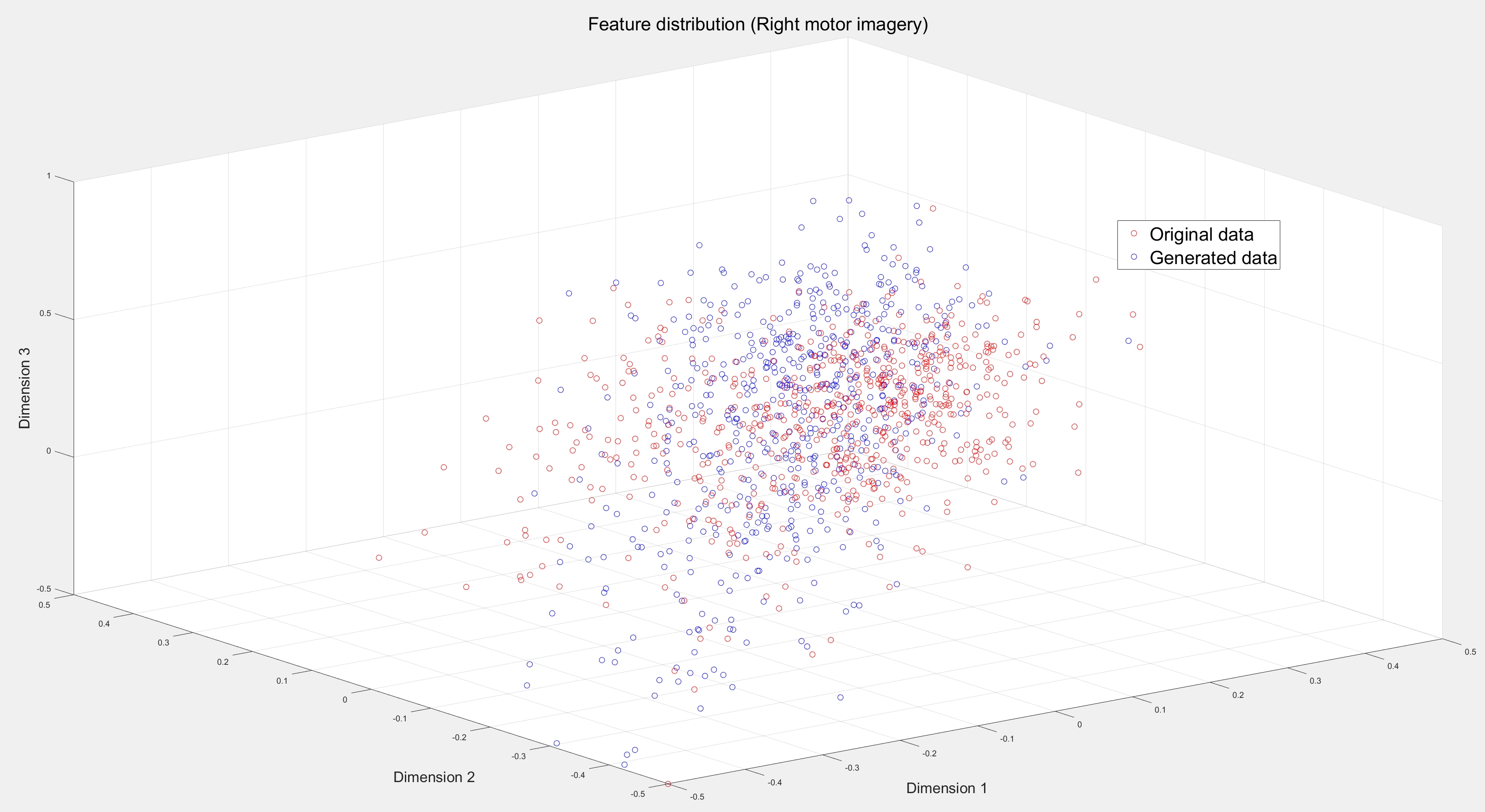}
  \caption{This illustration is a three-dimensional representation of the feature distribution for EEG signals (FID=0.0244), wherein the red data points denote the original data feature distribution associated with the motor imagery of the right hand from BCI competition 4-2a dataset, and the blue data points signify the feature distribution of the generated data corresponding to the motor imagery of the right hand from the same dataset.}
  \label{Distribution2}
\end{figure}

Employing an analogous methodology, we calculate the compressed features of the motor imagery for the synthesized data, followed by the creation of their respective feature distribution maps. Ultimately, all datasets are amalgamated, culminating in the production of a comprehensive feature distribution map, as illustrated in Fig.\ref{Distribution1} and Fig.\ref{Distribution2}.

The findings demonstrate that the difference in distribution between the synthesized signal and the original signal is less than 0.05, suggesting that the distribution of the images generated by the generative model closely resembles the distribution of authentic data. This suggests that the utilized generative model has robust learning capabilities, allowing it to effectively capture and replicate the statistical properties and intrinsic patterns of the original data. The EEG generative model independently selects an appropriate loss function to assess and direct the divergence between the synthesized data and the original dataset, ensuring the congruence of the distributional attributes between the generated and original data. Through extensive iterative optimization, the model has been fine-tuned to yield generated data that is consistent with the features of the original distribution. This alignment in feature distribution is evidence of the model's expertise in learning from the original dataset and its proficiency in producing data that mirrors the statistical nuances of the original EEG data.

\section{Conclusion}
This paper presents a revolutionary approach to EEG signal generation, employing a cutting-edge diffusion model enhanced by reinforcement learning techniques. Our methodology responds to the critical demand for extensive and varied datasets in EEG research and applications, especially within the realm of brain-computer interfaces. By producing synthetic EEG signals that replicate both the temporal dynamics and spectral intricacies of genuine brain activity, our model surmounts the constraints imposed by data scarcity, participant fatigue, and privacy issues associated with conventional data collection practices. The incorporation of reinforcement learning into our diffusion model represents a pivotal innovation, providing a dynamic and adaptive framework for optimizing the generative process. This strategy not only ensures the authenticity and diversity of synthetic signals but also bolsters the generalizability of machine learning models trained using this data. Our validation using the BCI Competition IV 2a dataset and our proprietary EEG data collection highlights the robustness and effectiveness of our methodology. The outcomes demonstrate a substantial advancement in EEG data augmentation. The diffusion generation methodology delineated in this paper represents a significant leap forward in the field of EEG signal synthesis. It provides a scalable, efficient, and privacy-preserving solution to the challenges of data acquisition, set to unlock new possibilities in brain-computer interface technologies and to contribute to the broader landscape of neuroscientific research and clinical practice.

\nocite{*}
\bibliographystyle{IEEEtran}
\bibliography{Reference}

\begin{thebibliography}{10}
\providecommand{\url}[1]{#1}
\csname url@samestyle\endcsname
\providecommand{\newblock}{\relax}
\providecommand{\bibinfo}[2]{#2}
\providecommand{\BIBentrySTDinterwordspacing}{\spaceskip=0pt\relax}
\providecommand{\BIBentryALTinterwordstretchfactor}{4}
\providecommand{\BIBentryALTinterwordspacing}{\spaceskip=\fontdimen2\font plus
\BIBentryALTinterwordstretchfactor\fontdimen3\font minus \fontdimen4\font\relax}
\providecommand{\BIBforeignlanguage}[2]{{%
\expandafter\ifx\csname l@#1\endcsname\relax
\typeout{** WARNING: IEEEtran.bst: No hyphenation pattern has been}%
\typeout{** loaded for the language `#1'. Using the pattern for}%
\typeout{** the default language instead.}%
\else
\language=\csname l@#1\endcsname
\fi
#2}}
\providecommand{\BIBdecl}{\relax}
\BIBdecl

\bibitem{Ref8}
\BIBentryALTinterwordspacing
Y.~Jiao, Y.~Deng, Y.~Luo, and B.-L. Lu, ``Driver sleepiness detection from eeg and eog signals using gan and lstm networks,'' \emph{Neurocomputing}, vol. 408, pp. 100--111, 2020. [Online]. Available: \url{https://www.sciencedirect.com/science/article/pii/S0925231220303325}
\BIBentrySTDinterwordspacing

\bibitem{Ref4}
J.~Tan, B.~Wu, and Y.~Ma, ``A visual characteristic land-scape design for eeg signal based on lstm-gan,'' \emph{IEEE Access}, vol.~12, pp. 41\,896--41\,907, 2024.

\bibitem{Ref6}
S.~Cohen, O.~Katz, D.~Presil, O.~Arbili, and L.~Rokach, ``Ensemble learning for alcoholism classification using eeg signals,'' \emph{IEEE Sensors Journal}, vol.~23, no.~15, pp. 17\,714--17\,724, 2023.

\bibitem{Ref10}
W.~Ko, E.~Jeon, and H.-I. Suk, ``A novel rl-assisted deep learning framework for task-informative signals selection and classification for spontaneous bcis,'' \emph{IEEE Transactions on Industrial Informatics}, vol.~18, no.~3, pp. 1873--1882, 2022.

\bibitem{Ref7}
\BIBentryALTinterwordspacing
J.~Kwon and C.-H. Im, ``Novel signal-to-signal translation method based on stargan to generate artificial eeg for ssvep-based brain-computer interfaces,'' \emph{Expert Systems with Applications}, vol. 203, p. 117574, 2022. [Online]. Available: \url{https://www.sciencedirect.com/science/article/pii/S0957417422008880}
\BIBentrySTDinterwordspacing

\bibitem{Ref11}
J.~Han, X.~Gu, G.-Z. Yang, and B.~Lo, ``Noise-factorized disentangled representation learning for generalizable motor imagery eeg classification,'' \emph{IEEE Journal of Biomedical and Health Informatics}, 2023.

\bibitem{Ref17}
H.~Zhang, H.~Ji, J.~Yu, J.~Li, L.~Jin, L.~Liu, Z.~Bai, and C.~Ye, ``Subject-independent eeg classification based on a hybrid neural network,'' \emph{Frontiers in Neuroscience}, vol.~17, p. 1124089, 2023.

\bibitem{Ref15}
\BIBentryALTinterwordspacing
J.~Yang, H.~Yu, T.~Shen, Y.~Song, and Z.~Chen, ``4-class mi-eeg signal generation and recognition with cvae-gan,'' \emph{Applied Sciences}, vol.~11, no.~4, 2021. [Online]. Available: \url{https://www.mdpi.com/2076-3417/11/4/1798}
\BIBentrySTDinterwordspacing

\bibitem{Ref16}
F.~Fahimi, S.~Dosen, K.~K. Ang, N.~Mrachacz-Kersting, and C.~Guan, ``Generative adversarial networks-based data augmentation for brain–computer interface,'' \emph{IEEE Transactions on Neural Networks and Learning Systems}, vol.~32, no.~9, pp. 4039--4051, 2021.

\bibitem{Ref9}
\BIBentryALTinterwordspacing
R.~Fu, Y.~Wang, and C.~Jia, ``A new data augmentation method for eeg features based on the hybrid model of broad-deep networks,'' \emph{Expert Systems with Applications}, vol. 202, p. 117386, 2022. [Online]. Available: \url{https://www.sciencedirect.com/science/article/pii/S0957417422007321}
\BIBentrySTDinterwordspacing

\bibitem{Ref14}
\BIBentryALTinterwordspacing
Q.~Liu, J.~Hao, and Y.~Guo, ``Eeg data augmentation for emotion recognition with a task-driven gan,'' \emph{Algorithms}, vol.~16, no.~2, 2023. [Online]. Available: \url{https://www.mdpi.com/1999-4893/16/2/118}
\BIBentrySTDinterwordspacing

\bibitem{Ref20}
W.~Qiao, L.~Sun, J.~Wu, P.~Wang, J.~Li, and M.~Zhao, ``Eeg emotion recognition model based on attention and gan,'' \emph{IEEE Access}, vol.~12, pp. 32\,308--32\,319, 2024.

\bibitem{Ref2}
\BIBentryALTinterwordspacing
Y.~Luo, L.-Z. Zhu, Z.-Y. Wan, and B.-L. Lu, ``Data augmentation for enhancing eeg-based emotion recognition with deep generative models,'' \emph{Journal of Neural Engineering}, vol.~17, no.~5, p. 056021, oct 2020. [Online]. Available: \url{https://dx.doi.org/10.1088/1741-2552/abb580}
\BIBentrySTDinterwordspacing

\bibitem{Ref13}
S.~S. Gilakjani and H.~Al~Osman, ``A graph neural network for eeg-based emotion recognition with contrastive learning and generative adversarial neural network data augmentation,'' \emph{IEEE Access}, 2023.

\bibitem{Ref5}
K.~Rasheed, J.~Qadir, T.~J. O’Brien, L.~Kuhlmann, and A.~Razi, ``A generative model to synthesize eeg data for epileptic seizure prediction,'' \emph{IEEE Transactions on Neural Systems and Rehabilitation Engineering}, vol.~29, pp. 2322--2332, 2021.

\bibitem{BCI4-2a}
C.~Brunner, R.~Leeb, G.~M{\"u}ller-Putz, A.~Schl{\"o}gl, and G.~Pfurtscheller, ``Bci competition 2008--graz data set a,'' \emph{Institute for knowledge discovery (laboratory of brain-computer interfaces), Graz University of Technology}, vol.~16, pp. 1--6, 2008.

\bibitem{CSP}
B.~Blankertz, R.~Tomioka, S.~Lemm, M.~Kawanabe, and K.-r. Muller, ``Optimizing spatial filters for robust eeg single-trial analysis,'' \emph{IEEE Signal Processing Magazine}, vol.~25, no.~1, pp. 41--56, 2008.

\bibitem{ResNet}
\BIBentryALTinterwordspacing
X.~Du, K.~Li, Y.~Lv, and S.~Qiu, ``Motor imaging eeg signal recognition of resnet18 network based on deformable convolution,'' \emph{Electronics}, vol.~11, no.~22, 2022. [Online]. Available: \url{https://www.mdpi.com/2079-9292/11/22/3674}
\BIBentrySTDinterwordspacing

\bibitem{CP-MixedNet}
\BIBentryALTinterwordspacing
D.~Nath, Anubhav, M.~Singh, D.~Sethia, D.~Kalra, and S.~Indu, ``A comparative study of subject-dependent and subject-independent strategies for eeg-based emotion recognition using lstm network,'' in \emph{Proceedings of the 2020 4th International Conference on Compute and Data Analysis}, ser. ICCDA '20.\hskip 1em plus 0.5em minus 0.4em\relax New York, NY, USA: Association for Computing Machinery, 2020, p. 142–147. [Online]. Available: \url{https://doi.org/10.1145/3388142.3388167}
\BIBentrySTDinterwordspacing

\bibitem{3DCNN}
X.~Zhao, H.~Zhang, G.~Zhu, F.~You, S.~Kuang, and L.~Sun, ``A multi-branch 3d convolutional neural network for eeg-based motor imagery classification,'' \emph{IEEE Transactions on Neural Systems and Rehabilitation Engineering}, vol.~27, no.~10, pp. 2164--2177, 2019.

\bibitem{EEGNet}
V.~J. Lawhern, A.~J. Solon, N.~R. Waytowich, S.~M. Gordon, C.~P. Hung, and B.~J. Lance, ``Eegnet: a compact convolutional neural network for eeg-based brain--computer interfaces,'' \emph{Journal of neural engineering}, vol.~15, no.~5, p. 056013, 2018.

\bibitem{Shallow}
Y.~Li, X.-R. Zhang, B.~Zhang, M.-Y. Lei, W.-G. Cui, and Y.-Z. Guo, ``A channel-projection mixed-scale convolutional neural network for motor imagery eeg decoding,'' \emph{IEEE Transactions on Neural Systems and Rehabilitation Engineering}, vol.~27, no.~6, pp. 1170--1180, 2019.

\bibitem{DAFBCSP}
Y.~An, H.~K. Lam, and S.~H. Ling, ``Multi-classification for eeg motor imagery signals using data evaluation-based auto-selected regularized fbcsp and convolutional neural network,'' \emph{Neural Computing and Applications}, vol.~35, no.~16, pp. 12\,001--12\,027, 2023.

\end{thebibliography}

\vfill

\end{document}